\documentclass{sig-alternate}
\usepackage{mathptmx} 
\usepackage{amsmath,amssymb,amsfonts,stmaryrd}
\usepackage{algorithmic}
\usepackage{fancyhdr}
\usepackage[normalem]{ulem}
\usepackage[hyphens]{url}
\usepackage[sort,nocompress]{cite}
\usepackage[final]{microtype}
\usepackage[keeplastbox]{flushend}
\usepackage[bookmarks=true,breaklinks=true,letterpaper=true,colorlinks,citecolor=blue,linkcolor=blue,urlcolor=blue]{hyperref}

\usepackage{mathrsfs}
\usepackage{eucal}
\usepackage{relsize}

\fancypagestyle{firstpage}{
  \fancyhf{}
  
  \fancyfoot[C]{\thepage}
}

\usepackage{caption}
\usepackage{subcaption}

\usepackage{authblk}

\usepackage{multirow}

\makeatletter
\newcommand*{\rom}[1]{\expandafter\@slowromancap\romannumeral #1@}
\makeatother

\title{A Scalable, Fast and Programmable Neural Decoder for Fault-Tolerant Quantum Computation Using Surface Codes}

\begin{document}

\author[ \thanks{Mengyu Zhang and Xiangyu Ren are joint first authors.} ]{Mengyu Zhang}
\author[ $^*$]{Xiangyu Ren}
\author[ ]{Guanglei Xi}
\author[ ]{Zhenxing Zhang}
\author[ ]{Qiaonian Yu}
\author[ ]{\\Fuming Liu}
\author[ ]{Hualiang Zhang}
\author[ $^\dag$]{Shengyu Zhang}
\author[ $^\dag$]{Yi-Cong Zheng}
\affil[ ]{Tencent Quantum Laboratory, Tencent, Shenzhen, Guangdong 518507, China}
\affil[$\dag$]{Corresponding authors: shengyzhang@tencent.com, yicongzheng@tencent.com}

\maketitle
\thispagestyle{firstpage}
\pagestyle{plain}

\begin{abstract}

Quantum error-correcting codes (QECCs) can eliminate the negative effects of quantum noise, the major obstacle to the execution of quantum algorithms. 
However, realizing practical quantum error correction (QEC) requires resolving many challenges to implement a high-performance real-time decoding system. Many decoding algorithms have been proposed and optimized in the past few decades, of which neural network (NNs) based solutions have drawn an increasing amount of attention due to their effectiveness and high efficiency.
Unfortunately, previous works on neural decoders are still at an early stage and have only relatively simple architectures, which makes them unsuitable for practical fault-tolerant quantum error correction (FTQEC).

In this work, we propose a scalable, low-latency and programmable neural decoding system to meet the requirements of FTQEC for rotated surface codes (RSC). Firstly, we propose a hardware-efficient NN decoding algorithm with relatively low complexity and high accuracy. Secondly, we develop a customized decoder architecture for our algorithm and carry out architectural optimizations to reduce decoding latency. 
Thirdly, our proposed programmable architecture boosts the scalability and flexibility of the decoder by maximizing parallelism. Fourthly, we build an FPGA-based decoding system with integrated control hardware to comprehensively evaluate our design.
Our $L=5$ ($L$ is the code distance) decoder achieves an extremely low decoding latency of 197 ns, and the $L=7$ configuration also requires only 1.136 $\mu$s, both taking $2L$ rounds of syndrome measurements as input. The accuracy results of our system are close to minimum weight perfect matching (MWPM). Furthermore, our programmable architecture reduces hardware resource consumption by up to $3.0\times$ with only a small latency loss. We validated our approach in real-world scenarios by conducting a proof-of-concept benchmark with practical noise models, including one derived from experimental data gathered from physical hardware.




\end{abstract}

\section{Introduction}\label{sec:intro}
Quantum computers offer a tremendous computational advantage on numerous important problems, but qubits are fragile and easily affected by noises that deteriorate computation fidelity quickly. Quantum error-correcting codes (QECCs) and the theory of fault-tolerant quantum computation (FTQC) are backbones for large-scale quantum computation. FTQC can perform operations at any scale and obtain reliable results on error-prone quantum hardware, as long as noise strength is under a certain threshold~\cite{Shor:1996:56,Kitaev:2003:2,aharonov2006fault,Aliferis:2006:97,QECbook:2013}.
The number of qubits on a single chip has been rapidly increasing~\cite{bravyi2022future,IBM_roadmap}, but the realization of fault-tolerant quantum error-correcting (FTQEC) schemes is still challenging and has not yet been surmounted. FTQEC introduces redundant resources to encode information into code space and decode them after computation. Among various QECCs proposed in previous 2-3 decades, surface codes~\cite{Kitaev:2003:2,bravyi1998quantum,Dennis:2002:4452,Folwer2012PhysRevA.86.032324} are considered the most promising scheme for solid-state platforms, as they require only nearest-neighbor operations.

\begin{figure}[t]
\centering
\includegraphics[width=0.46\textwidth]{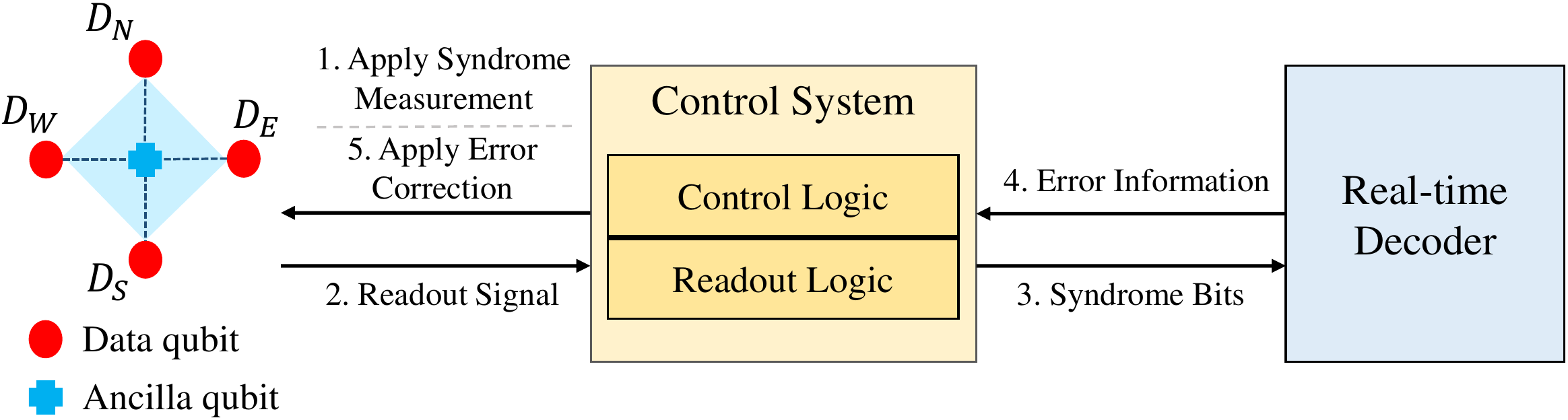}
\setlength{\abovecaptionskip}{3pt}
\captionsetup{labelfont={small,bf}, font={small,bf}}
\caption{
Steps required for QEC after logical qubit encoding. 
}
\label{fig:qec_steps}
\vspace{-6mm}
\end{figure}

The process of FTQEC based on surface code is shown in Figure~\ref{fig:qec_steps}. A logical qubit is encoded on multiple data qubits, interspersed (also see later Figure~\ref{fig:RSC}) with ancilla qubits which are used for performing multiple rounds of syndrome measurements (SM) to collect sufficient error information without destroying the state of data qubits.
A control system consisting of control and readout logic applies syndrome measurement signals and discriminates the returned results. 
The collected syndrome bits are then transferred to the real-time decoder and analyzed to determine the exact locations and types of the errors \emph{in-situ}. Finally, the control logic applies corresponding error correction signals to the data qubits to complete a QECC cycle.

Many challenges rise in designing and implementing good decoders. The most prominent ones are believed to be: \textbf{(1) High-performance.} The decoding algorithm should reduce the logical error rate as much as possible. Since QECCs cost many extra qubits, their error correction capacity should be fully explored to get paid off. 
\textbf{(2) Scalability.} The decoding algorithms should be intrinsically parallelizable so that their hardware implementation can scale up with the code distance more efficiently by fully utilizing computational resources.
On this basis, it is also necessary to perform hardware architectural optimizations to alleviate the high resource consumption caused by the growing size of the FTQC.
\textbf{(3) Low-latency.} The decoding algorithms need to be executed fast enough to avoid error accumulation. More specifically, the latency for the whole FTQEC process should be short to catch up with syndrome generation so that one can \emph{physically} correct and control data qubits before non-Clifford gates~\cite{zheng2020constant,riesebos2017pauli}. Failure to achieve this constraint will lead to backlog problem~\cite{terhal2015quantum,holmes2020nisq+,chamberland2022techniques,skoric2022parallel}, which causes exponential computation overhead to kill any quantum advantage. For state-of-the-art superconducting qubits with lifetime 150-300 $\mu$s~\cite{Haohua2022ExperimentalQA}, \emph{FTQEC within 1.5 $\mu$s is highly preferred}. 
\textbf{(4) Flexibility.} Decoders need to work in lots of different scenarios with various noise levels, code distances, code deformations~\cite{fowler2009high,Folwer2012PhysRevA.86.032324} and lattice surgery~\cite{horsman2012surface_lattice_surgery,ueno2022qulatis,ueno2022neo} suitable for FT operations. Decoders that can be programmed to switch between different scenarios would significantly broaden the applicability. 

In addition to these challenges, the implementation of FTQEC is a system-level task---the decoder has to be seamlessly integrated into the control system to be fully functional.



A recent review \cite{battistel2023real} discusses a range of candidates for real-time error decoding. Among them are minimum weight perfect matching~(MWPM)~\cite{Edmonds:1965:449,wang2011surface,fowler2012towards} and Union-Find~(UF)~\cite{Union_find_delfosse2017almost,weighted_UF_huang2020,delfosse2020hierarchical}. MWPM is the most well-known and advanced, but suffers from being too complicated. Indeed, its complexity scales as $O(L^{9})$ ($L$ is the distance of the code). Even after tremendous optimization~\cite{fowler2012towardsclassical_prl,fowler2012towards,fowler2015o(1),higgott2023sparse}, it has yet to illustrate its low-latency decoding on real devices even for small $L$. UF has reasonably good decoding performance, with complexity almost proportional to $L^3$. Both algorithms can be directly deployed through Look-Up Table~(LUT) solution~\cite{das2021lilliput}, but is difficult to scale up since the number of entries grows exponentially with $L^3$ in both cases. UF hardware decoders have been proposed \cite{das2022afs, liyanage2023scalable}, but their actual performance is only evaluated under the phenomenological noise model, while incorporating complete noise would significantly slow the decoder.



Recently, neural networks (NNs) based solutions have attracted an increasing amount of attention ~\cite{Boltzman2018,savvas2017decoding, NNgeneral_framework2018, chamberland2018deep,  savvas2019comparing, savvas2020decoding, NN_colorcode2019,ni2020neural,gicev2021scalable,meinerz2022scalable,chamberland2022techniques,ueno2022neo} due to their high accuracy and computational efficiency. 
Previous works~\cite{overwater2022neural,chamberland2018deep, chamberland2022techniques} designed various neural decoders and analyzed their cost and performance for different hardware platforms. Despite the effectiveness in the reported settings, the algorithms and microarchitecture there are relatively primitive and may fail to fit real experimental environments due to their high latency or incomplete noise model. 
Moreover, to our knowledge, no solution regarding flexibility has been proposed in these prior works.
Consequently, the actual performance and latency of the \textit{entire decoding system} that can comprehensively address the above challenges has yet to be demonstrated.



To address these challenges, we propose a scalable, low-latency and programmable neural decoding system. The proposed neural network-based decoding algorithm has high performance and is customized for hardware-efficient deployment. 
Additionally, we present a decoder microarchitecture design that optimizes the resource allocation and exploits parallelism in multiple rounds of SMs for low latency.
To comprehensively evaluate the performance of the proposed system, we implement a field-programmable gate arrays (FPGAs)-based decoding system, including the decoder as well as other control hardware. 
To demonstrate the effectiveness of our solution, we use a \textit{circuit-level noise model}, where noises due to imperfect qubits, gates, and measurements are all considered.

The assessment indicates that our decoder's accuracy at $L=5$, amassing ten rounds of SM results, approximates MWPM. However, the decoding latency is experimentally ascertained at 197 ns, substantially quicker than MWPM on CPUs~\cite{fowler2015o(1),higgott2022pymatching, higgott2023sparse}.

Furthermore, we employed a noise model derived from experimental data obtained from the Google QEC study to train and test our decoder\cite{google2023suppressing,Google_data}, proving our solution is practical in real-world environments.


In contrast to conventional NN accelerators, which emphasize average throughput and avoid using resources simultaneously for single-task latency reduction, quantum error decoding needs to maximize resource utilization within a specific time. We then propose a programmable architecture to exploit this feature. This design reuses general-purpose arithmetic units for diverse decoding configurations, efficiently employing computational resources to minimize latency, enhancing scalability, and addressing flexibility challenges.

Overall, our contributions in this work are:
\begin{enumerate}
    \vspace{-2mm}
    \item We present an innovative, efficient fault-tolerant neural decoding algorithm based on stepper 3D CNN~\cite{ji2012_3dcnn} and multi-task learning~\cite{caruna1993multitask}. It exhibits competitive accuracy compared to MWPM, while significantly reducing latency. Its NN layer count scales as $O(\log L)$, rendering it scalable for future applications requiring large $L$ and minimal latency. Moreover, the computational complexity scales a $O\left(L^3\right)$, which is comparable to UF and more conducive to hardware implementation.
    \vspace{-2mm}

    \item We introduce a decoder microarchitecture optimized for achieving low latency while preserving high accuracy. Our FPGA-based implementations for $L=5$ and $L=7$ attain decoding latencies of \textbf{197 ns} and \textbf{1.136 $\mu$s}, respectively. Both configurations incorporate $2L$ rounds of syndrome measurements.
    \vspace{-2mm}
    \item We build a complete decoding system that integrates our decoder and customized control hardware, achieving an overall system latency of \textbf{540 ns}. This system is the fastest real-time fault-tolerant decoding system ever built and testified for dozens of qubits surface code.
    \vspace{-2mm}
    \item We develop a programmable architecture to accommodate diverse decoding configurations with flexibility. In comparison to traditional approaches, our design maximizes hardware resource utilization and diminishes resource overhead by up to \textbf{3.0$\times$}, incurring only a minimal latency expense. Additionally, the ASIC implementation of our programmable architecture is compatible with diverse decoder configurations, encompassing distinct network structures and code distances.
\end{enumerate}
\section{Preliminaries and Motivation}

\subsection{Rotated surface code }

\begin{figure}[t]
\centering
\includegraphics[width=0.48\textwidth]{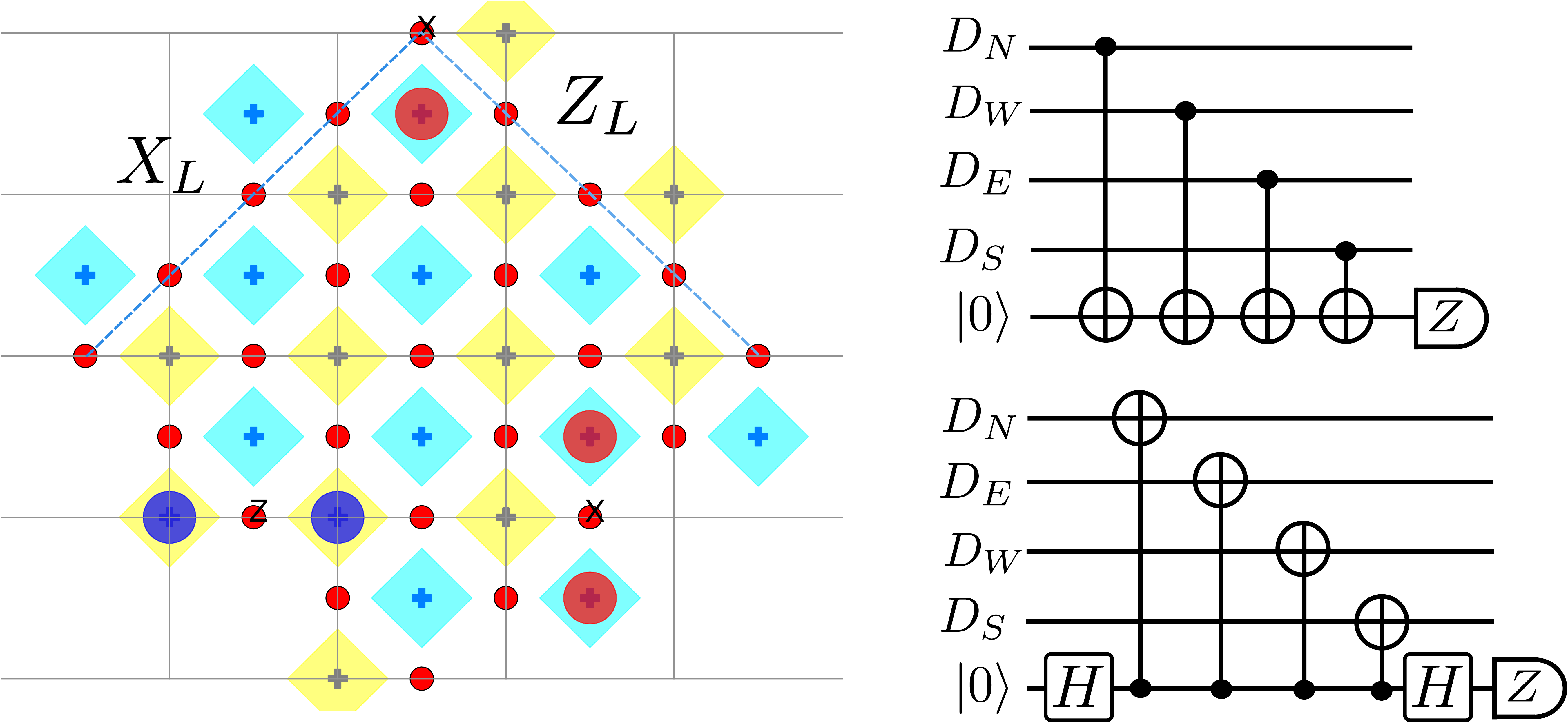}
\setlength{\abovecaptionskip}{3pt}
\captionsetup{labelfont={small,bf}, font={small,bf}}
\caption{
(left) RSC with $L=5$ with 25 data qubits (red dots) encoding 1 logical qubits characterized by a particular choices of the logical operator $X_L$ and $Z_L$ (dashed lines). $Z_p$ and $X_v$ are indicated as cyan and yellow plaquettes, respectively. Ancilla qubits (crosses) for $Z_p$ and $X_v$ measurements are located at the plaquettes and vertices. Several data qubits are affected by Pauli errors. Measuring the $Z_p$s and $X_v$s yields $1$-valued syndrome bits of certain $X_v$ (dark blue) and $Z_p$ operators (red). (right) A single round of SM circuits for $Z_p$ and $X_v$.
\label{fig:RSC}}
\vspace{-5mm}
\end{figure}

Surface codes are a family of stabilizer codes defined on a 2D square lattice. The smallest version of planar surface codes,
which requires the least amount of physical qubits, are known as
the \textit{rotated surface codes}~(RSC). In this paper, we focus on the RSC consisting of $L\times L$ data qubits, as shown in Figure ~\ref{fig:RSC} for $L=5$. The stabilizer generators of surface codes are two different kinds of operators: $X_v=\bigotimes_{i\in v}X_i$ and $Z_p = \bigotimes_{i\in p}Z_i,$ that represent vertices ($X_v$, or $X$ type) and plaquette ($Z_p$, or $Z$ type) on the square lattice. For each $v$ (ancillary qubit in yellow plaquette), $X_v$ is the tensor product of $X$ operators on the four red qubits around the yellow plaquette; similarly for each $Z_p$ in the cyan plaquette. The operators $X_v$ and $Z_p$ generate the stabilizer group $\mathcal{S}$. If no error of any kind occurs, the syndrome bits are all 0. If $X$ or $Z$ errors occur, the syndrome bits of the stabilizer generators that anti-commute with errors will be flipped to $1$. Each $X_v$ or $Z_p$ needs an extra ancillary qubit to interact with the data qubits around it in a specific order for syndrome measurements~(SM). See Figure.~\ref{fig:RSC} for an example of errors as well as SM circuits to extract the syndrome bits.  All equivalent logical operators form a topology class, called the \textit{homology class}, which is also the logical class for surface code. For each homology class $\mathcal{L}$, we choose 
a representative $\mathcal{L}_c$ which has the minimum weight $L$ in $\mathcal L$. This weight is defined as the distance of RSC. It is known that arbitrary errors on any $\lfloor\frac{L-1}{2}\rfloor$ qubits can be corrected. 
If too many errors occur, 
the decoding algorithm fails to correct the errors, which causes failure of computation.

RSCs are greatly favored in solid-state platforms due to their low requirement on the number of physical qubits and connections between them. Recent experimental progresses of superconducting platforms have enabled the realization of RSC encoded states using off-line decoding based on multiple rounds of SM~\cite{andersen2020repeated,google_surface_code_1,google2023suppressing,ustc_surface_code,eth_krinner2022realizing}.

\subsection{FTQC and real-time decoding}
Quantum noises occur at all  places during the computation. One needs to apply SM circuits periodically to extract syndrome bits during the whole procedure of computation. 

The SM circuits need to be executed for all $X_v$ and $Z_p$ operators simultaneously. Note that the SM circuits themselves also suffer from gate and measurement noises, and the CNOT gates in SMs may propagate single-qubit error to two data qubits.  
To mitigate the effect caused by such propagation, the order of CNOTs acting on data qubits around ancilla should respect the distribution of logical operators~\cite{tomita2014low}---it maintains the alignment of the last two qubits involved with SM circuits so that they are perpendicular to the direction of the corresponding logical operators. Such alignment can reduce the effect of error propagation caused by SMs. 

In general, measuring syndromes once cannot distinguish errors on data qubits from measurement errors, which will quickly cause logical errors. Fortunately, with a sufficiently large number $T$ of rounds of SM, one can establish reliable syndrome information for FTQEC. 
\begin{figure}[t]
\centering
\includegraphics[width=0.42\textwidth]{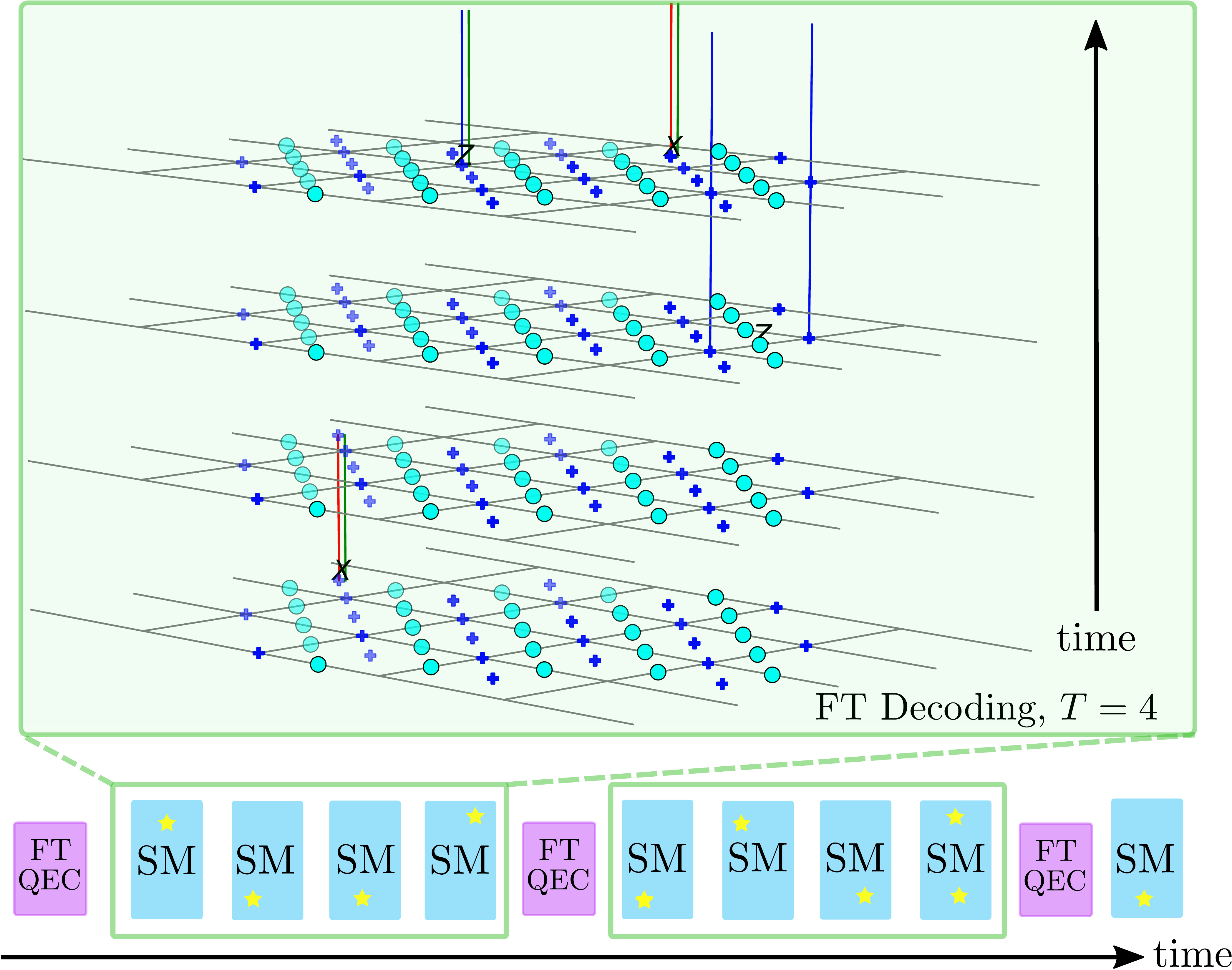}
\setlength{\abovecaptionskip}{3pt}
\captionsetup{labelfont={small,bf}, font={small,bf}}
\caption{
An illustration of repeated real-time FTQEC every 4 rounds of SMs. The effective data and measurement errors caused by a realization of circuit-level noise are shown in space-time. The red (blue) lines are syndrome history of $X_v$s~($Z_p$s). The green line represents the history of measurement errors. The FTQEC is applied every $T$ rounds of SMs and the correction is applied  on the data qubits right after the decoding.
\label{fig:3d_RTFTQC}}
\vspace{-6mm}
\end{figure}

Non-Clifford~(like the logical {\rm T} gate) gates bring more challenges. If only Clifford gates exist, 
the decoding can be postponed to the end of storage by post-processing all the syndrome bits in the space-time history following the Pauli frame change. 
However, quantum computational advantage does need non-Clifford gates~\cite{gottesman1997}, and when they exist, the SMs after them introduce random Pauli frames and destroy the historical error information. To resolve this, all errors must be corrected before non-Clifford gates. 
This brings a real-time constraint for the decoding and error correction: after every $T\sim O(L)$~\cite{Dennis:2002:4452} rounds of SMs, the FT decoder takes these $T$ slides of syndrome bits as input to infer the most likely errors on the data qubits; these errors then need to be corrected before next rounds of gate operations. Such a procedure needs to be finished at a speed faster than SMs to avoid backlogs problem which causes exponential computation time overhead~\cite{holmes2020nisq+,chamberland2022techniques,terhal2015quantum}.
The illustration of repeated real-time FTQEC is shown in Figure~\ref{fig:3d_RTFTQC} for $T=4$. 

\subsection{Motivation: FTQEC for Near-term and Large Scale}

Previous work has shown successful execution of real-time FTQEC based on 3-qubit repetition code~\cite{BBNriste2020real} recently, but only $X$ (or $Z$) errors can be corrected. Some state-of-the-art superconducting quantum hardware demonstrated the implementation of an RSC with $L=5$ with offline decoding. Real-time FTQEC is expected to be achieved in the coming years. To that end, building real-time decoding systems for $L=5$ and beyond based on off-the-shelf devices such as FPGAs is a major goal in the near term.

In the long term, problems like integer factorization or quantum simulation with FTQC require hundreds or thousands of logical qubits and millions of circuit layers. To achieve this, it is essential to minimize the hardware resource costs in designing large-scale high-performance decoders, especially when considering the future use of emerging technologies such as cryo-electronics.



\section{Evaluation Methodology}

\subsection{Noise Model}\label{sec:noise_model}

We use \emph{circuit-level Pauli noise} for our evaluation: assume that during each SM, each data qubit undergoes an $X$, $Y$, or $Z$ error each with probability $p_s/3$, called the \textit{storage noise}. For CNOTs, noises are modeled as perfect gates followed by one of the 15 possible two-qubit Pauli operators, with equal probability $p_g/15$, which is called the \textit{gate noise}. The measurement of a single physical qubit suffers a classical bit-flip error with probability $p_m$, called \textit{measurement noise}. Recent experiments~\cite{arute2019quantum,google_surface_code_1} show that it can catch the essence of practical noises process to a great extent.

The \emph{phenomenological noise model}, employed extensively in prior research, does not account for gate noise. It is crucial to acknowledge that incorporating CNOT errors results in a considerably more computationally demanding decoding process, increased latency, and diminished accuracy. To illustrate the difference, we collected the probability distribution of Hamming weights (HW) of syndrome bits under these two noise models. We generated one million samples and the results are shown in Table \ref{table:Hamming weights}.

\begin{scriptsize}
\begin{table}[h!]
  \small
  \centering
  \begin{tabular}{|c|c|c|c|}
    \hline
    HW ($L = 5$, $T = 10$, & Probability & HW ($L = 7$, $T = 14$, & Probability\\
    circuit-level) & & circuit-level) & \\
    \hline
    23 & 1.62e-4 & 50 & 1.12e-4 \\
    \hline
    24 & 8.4e-5 & 51 & 6.8e-5 \\
    \hline
    \hline
    HW ($L = 5$, $T = 10$, & Probability & HW ($L = 7$, $T = 14$, & Probability \\
    phenomenological) & & phenomenological) & \\
    \hline
    15 & 4.9e-5 & 27 & 4.4e-5 \\
    \hline
    16 & 2.4e-5 & 28 & 2.6e-5 \\
    \hline
  \end{tabular}
  \setlength{\abovecaptionskip}{5pt}
  \captionsetup{labelfont={small,bf}, font={small,bf}}
  \caption{Hamming weights sampled at $p=0.006$ for different configurations when the probability decays to 0.}
  \label{table:Hamming weights}
  \vspace{-4mm}
\end{table}
\end{scriptsize}


It is clear that the Hamming weight of the syndromes array undergoes a marked reduction when moving from the circuit-level noise model to the phenomenological model. Consequently, we contend that employing a more comprehensive noise model is essential, as it aids in assessing the applicability of the decoder design for real-world experiments, while simultaneously introducing more challenges in decoding.

Moreover, we also test our decoder based on an effective circuit-level noise model extracted from Google's experiments on 72-qubit Sycamore device~\cite{google2023suppressing, Google_data}. This model can be employed to generate training data for our NN algorithm, so that we can test the practicality of our solution in realistic environments. 

\subsection{Evaluation Framework}

We used Monte Carlo simulation for system verification and built an hardware platform (including decoder and other control hardware) to evaluate the actual performance of the decoding system following the procedure of Figure~\ref{fig:qec_steps}.
The error is assigned for SMs according to the noise model in software to sample syndrome bits. These bits are then translated into waveform data using a set of demodulation and thresholding parameters, which is also configured in the readout module. This procedure mimics the readout and signal processing in actual experiments. Finally, they are transmitted to the decoder for error correction. The process repeats for each trail trajectory until a decoding failure occurs, and average time duration $\bar{\tau}$ is recorded. The logical error rate is defined as $1/(T \bar{\tau})$. At least 400 such trajectories are carried out for each physical error rate to calculate the logical error rate. With this platform, we evaluate the entire decoding process on classical hardware. The implementation of this framework is introduced in Section \ref{sec:implementation}.


\subsection{Target Hardware Platform}

Regarding the near-term goal, we focus on FPGAs, which can be easily integrate into existing centralized control systems \cite{fu2017experimental, zhang2021exploiting} and accomodated to the frequent updates of early-stage experimental set-ups.
The use of ASICs becomes a natural choice as the system size further grows to future large-scale FTQCs. 
Emerging technologies such as cryo-CMOS put forward higher requirements for power budget and other metrics. Although these limitations are not discussed in detail in this work, resource efficiency and higher scalability presented in our decoder can help alleviate these issues.
In this work, we demonstrate the performance of our decoding system with a complete FPGA-based implementation. FPGAs are also used to evaluate the scalability and flexibility of our decoder in large-scale FTQEC scenarios. Our solution can be easily extended to ASICs when required.

FTQC requires RSCs with at least $L\geq3$ to correct both $X$ and $Z$ errors. The smallest case of $L=3$ can be implemented directly through LUTs because of the small number of syndrome bits. 
Therefore, we focus on the case of $L=5$ and $L=7$ when studying near-term error decoding, and $L>7$ for future large-scale FTQEC.

\subsection{Syndrome Measurement Rounds}

To ensure fault-tolerance validity, it is theoretically required that the number of syndrome measurement rounds ($T$) be equal to or greater than the code distance ($L$) \cite{dennis2002topological, fowler2012proof}, which is a common practice in previous error decoding research.
In the mean time, for $T$ larger than $2L$, the decoding complexity increases but has minimal effect on further lowering the logical error rate. Therefore, the number of SM rounds we choose in the evaluation is between $L$ and $2L$.

\section{FT neural decoding algorithm}\label{sec:NNdecoding_algorithm}
\subsection{Elementary Nueral Network}
An NN is a directed graph consists of multiple layers of nodes called \textit{neurons}. Each node $v$ is assigned a value $y_v$ and a \textit{bias} parameter $b_v$, and each edge $(p,v)$ is assigned a \textit{weight} parameter $W_{vp}$.
The value $y_v$ is obtained from applying an activation function $\mathscr{A}$ to the summation of the bias $b_v$ and the $W_{vp}$-weighted sum of the values $y_p$ of the incoming neighbor nodes $p$:

\begin{equation}\label{eq:FFN}
y_v = \mathscr{A} (\sum_{p\rightarrow v}W_{vp}y_p + {b}_v ).
\end{equation}
It should be easy to compute the derivative of the activation function $\mathscr{A}$. Common choices of $\mathscr{A}$ include sigmoid, $\mathrm{Tahn}$ and rectified linear unit~($\mathrm{ReLU}$) and $\mathrm{Leaky ReLU}$, the latter two of which are used in this work. One can also apply an extra $\mathrm{Softmax}$ function on the values of the output neurons to generate a normalized output that can represent a distribution. 

The elementary NNs used in this paper are restricted to fully connected networks (FCN) and 3D convolutional NN (3D CNN)~\cite{ji2012_3dcnn}. 
These modules are chosen because of their good representation power to extract the important local features, as well as their simplicity to implement with digital circuits. 

\begin{figure}[ht]
\centering
\includegraphics[width=0.47\textwidth]{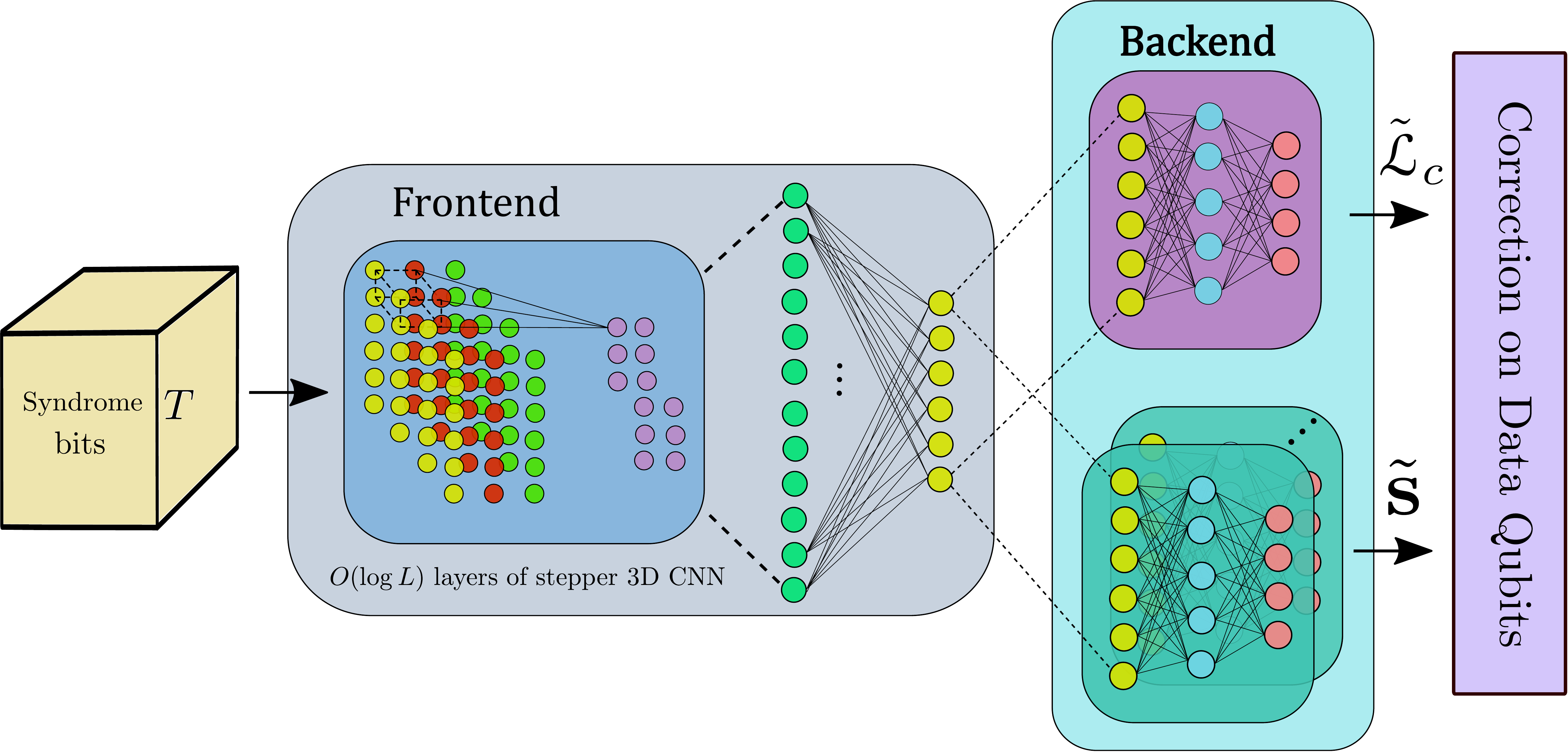}
\setlength{\abovecaptionskip}{3pt}
\captionsetup{labelfont={small,bf}, font={small,bf}}
\caption{A structure of FT neural decoding algorithm for RSC.
\label{fig:mtl_decoder}}
\vspace{-5mm}
\end{figure}

\subsection{Decoding on marginal posterior distribution}
The decoding algorithm can be viewed as a process of mapping the collected syndromes to $L^2$-fold Pauli operators. The $L^2$-fold Pauli group can be divided into $2^{L^2+1}$ classes:
\begin{equation}
\mathcal{C}_{\mathcal{L}_c, \mathbf{s}} = \{g\mathcal{L}_c\mathscr{T}(\mathbf{s})\ | \ g\in \mathcal{S}\}, \quad \mathbf{s}\in \mathbb{Z}_2^{L^2-1},
\end{equation}
where the elements in each class are equivalent with respect to RSC, and their representative are $\mathcal{L}_c\mathscr{T}(\mathbf{s})$. Here $\mathscr{T}(\mathbf{s})$ is the pure error given $\mathbf{s}$, which can directly calculated through an LUT~\cite{Poulin:2006:052333}. In this setting, the optimal way to infer the error on data qubits after $T$ rounds of SM from a measured $T\times (L+1)^2$ syndrome array $\mathbf{S}$ is:
\begin{equation}\label{eq:MAP}
\tilde{\mathcal{C}} = \arg\max_{\mathcal{L}_{{c}},\mathbf{s}}\Pr(\mathcal{C}_{\mathcal{L}_c, \mathbf{s}} | \ \mathbf{S})=\arg\max_{\mathcal{L}_{{c}},\mathbf{s}}\sum_{g\in \mathcal{S}} \Pr( g\mathcal{L}_c\mathscr{T}(\mathbf{s})|\ \mathbf{S})
\end{equation}
which can be recognized as a Maximum a \emph{Posteriori} (MAP) estimation. The distribution is over $2^{L^2+1}$ possible entries, which is intractable in general. To solve this, we decompose the binary string $\mathbf{s}$ into $m$ pieces:
$\mathbf{s} = \mathbf{s}_{1} \sqcup \mathbf{s}_{2} \cdots  \sqcup \mathbf{s}_{m}$, with $\sqcup$ being concatenation and $|\mathbf{s}_j|\sim O(1)$ for all $j$. We approximate Equation~$\eqref{eq:MAP}$ by the marginal posterior distribution:
\begin{equation*}
\tilde{E}=\arg\max_{\mathcal{L}_{{c}}}\sum_{g\in \mathcal{S}} \Pr(g \mathcal{L}_{c}  | \mathbf{S})\mathscr{T}\left(\bigsqcup_{j=1}^{m}\arg\max_{\mathbf{s}_{j}}\Pr(\mathbf{s}_{j}  | \mathbf{S})\right).
\end{equation*}
Such simplification neglects the correlation between different $\mathbf{s}_j$ of the optimal solution, which is a reasonable assumption since $\mathscr{T}(\mathbf{s}_i)$ and $\mathscr{T}(\mathbf{s}_i')$ are typically highly different operators even when the weight of $(\mathbf{s}_i\oplus\mathbf{s}_i')$ is small.     

\begin{figure*}[t]
\centering
\includegraphics[width=0.9\textwidth]{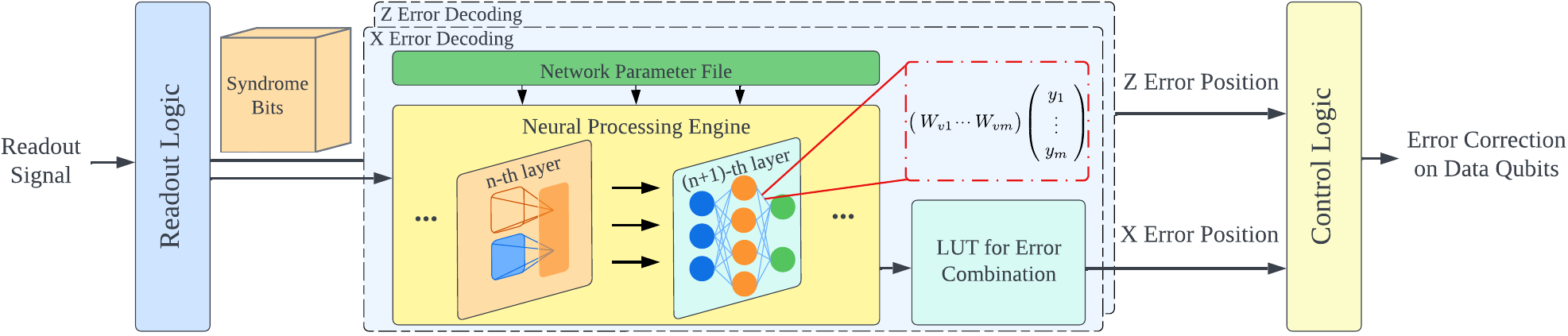}
\setlength{\abovecaptionskip}{3pt}
\captionsetup{labelfont={small,bf}, font={small,bf}}
\caption{Decoder overview.}
\label{fig:overview}
\vspace{-5mm}
\end{figure*}

\subsection{Multi-task learning neural decoder}
{We first introduce an end-to-end NN (see Figure~\ref{fig:mtl_decoder}) to simultaneously learn multiple marginal posterior distributions~\cite{caruna1993multitask}}. We separate the NN into the frontend and the backend parts. The frontend {consists of} multiple layers of 3D CNNs {followed by one layer of FCN} to extract common features. The input and output layers of 3D CNNs are two groups of 3D neuron arrays carrying feature information. Due to the space-time locality of $\mathbf{S}$, we assume that for each 3D neuron array, the correlation of the values of different neurons decays quickly with their distance. Hence, we implemented 3D CNNs in a stepper manner: their strides are roughly the same as the kernel sizes, which are bounded by some constant $K$, and the mappings focus on extracting local features. Since the sizes of 3D neuron arrays of the $i$-th layer shrink exponentially with $i$, both training and inference time of NNs do not increase much with the depth of 3D CNN part. 

The backend consists of $m+1$ multi-layer FCNs  
to approximate the marginal posterior distributions for $\mathcal{L}_c$ and {$\{\mathbf{s}_1, \ldots, \mathbf{s}_m\}$}. These multi-layer FCNs 
share the same input from the frontend, {which is trained to} 
extract sufficient features to calculate all the marginal posterior distributions. 

We use the sum of $\mathrm{CrossEntropy}$ for the output distributions as the loss function, and SGD/ADAM~\cite{Kingma2015AdamAM} for training. 
This Multi-task learning neural decoder (MTLND) is split into two NNs, to infer $X$($Z$) errors based solely on $Z$($X$) syndrome bits.

\subsection{Complexity analysis}
The computation elements for NNs here are exclusively multiplication and addition. With a stepper manner implementation of all 3D CNNs, the total number of layers in frontend is around $O(\log L)$. The sizes of all FCNs are chosen to be independent of $L$, with depth $O(1)$. Hence, the depth of the NNs is $O(\log L)$, which puts a small lower bound of computation latency if all layers can be sufficiently parallelized to finish in $O(1)$ steps. Suppose the kernel size is lower bounded by $k$. The total number of multiplication operations, which dominates the computation, is bounded from above by
\begin{equation}\label{eq:compu.comple}
C^2\sum_{i=1}^{\lceil \log L \rceil} K^3\frac{L^3}{k^{3i}}+ D \left(\frac{L^2}{\min\{|\mathbf{s}_j|\}}+2\right) \sim O(L^{3}),
\end{equation}
where $C$ and $D$ are the maximum number of input/output channels of the 3D CNN and of edges of each multi-layer FCN, respectively. Such complexity is competitive with UF. The total number of the parameters for each NNs can be bounded by:
\begin{equation}
C^2K^3\lceil\log_k(L)\rceil + D \left(\frac{L^2}{\min\{|\mathbf{s}_j|\}}+2\right)\sim O(L^2).
\end{equation}
This relatively slow scaling makes the hardware implementation feasible for loading  all the parameters into on-chip memories, whose sizes are often limited.

\subsection{Training and Quantization}
\noindent \textbf{Training}
The training data set is generated by simulating circuit-level noises at $p_s=p_g=p_m\sim0.006$---each sampled 3D syndrome $\mathbf{S}$ pairs with label $(\mathcal{L}_c,\mathbf{s})$. 
For $X$ ($Z$) errors, one may utilize either $Z$ ($X$) type syndromes or a combination of both $X$ and $Z$ syndromes as input for the MTLND. The latter approach offers superior accuracy but requires a significantly more intricate neural network structure.
The training is carried out through ADAM in Pytorch 1.5 with batch size 700-1000 for 8 to 10 epochs on two NVIDIA V100 GPUs. 

\noindent \textbf{Quantization}
We choose the non-saturating quantization scheme for all weights and biases~\cite{jacob2018quantization}. The outputs of each layer are re-scaled so that the input data of its consequent layer is maintained to be signed 8-bit integers. As we will see, it simplifies the implementation of arithmetic modules and data files, while incurring only small loss of accuracy.

\section{Decoder Overview}\label{sec:decoder_overview}






\subsection{Decoder Microarchitecture: A Big Picture}\label{sec:decoder_arch_overview}

Figure \ref{fig:overview} shows the microarchitecture of our proposed decoder. We describe and explain the main components and functions of the decoder as follows:


\noindent \textbf{Syndrome Bits.} Syndrome bits are measurement results obtained from classical readout logic. For RSCs with distance $L$, $T\sim O(L)$ rounds of measurements are required to guarantee fault tolerance. Better decoding accuracy requires larger $T$. These $T$ slices of syndrome bits are combined into a 3D array and fed into either $X$-type or $Z$-type decoding logic, depending on the ancilla type.

\noindent \textbf{Network Parameter File.} NN parameters are obtained offline through the training phase and loaded to the Network Parameter File before a quantum computation starts. Different sets of NN parameters need to be fetched during the decoding, demanding fast switching of various sets of parameters during real-time decoding. Therefore, we need to use on-chip memory to implement this module to avoid extensive memory loading delays. The entire storage is divided into two parts according to different data structures, one for storing weight matrices and the other for bias vectors. These parameters are originally floating-point numbers, which lead to complicated multiplications and large storage space. 
To improve the storage and computational efficiency, the parameters are quantized to 8-bit signed fixed-point numbers.


\noindent \textbf{Neural Processing Engine (NPE).}
This engine consists of the arithmetic units~(AUs) for NN computation. The operators allowed include 3D CNNs and FCNs, both of which involves repeated computation of vector inner products as in Equation~\eqref{eq:FFN}.
The multiplication-addition operations in  Equation \eqref{eq:FFN} take up the majority of computing resources in NPE. 
Since the bias vectors are accessed only once per iteration, they can also be stored in a series of simple registers.

\noindent \textbf{LUT for Error Combination.}
The error locations are identified and combined in this module.
For either $X$-type or $Z$-type error decoding, NPE generates one logical operators $\tilde{\mathcal{L}}_c^{X|Z}$ and $\frac{L^2-1}{2}$ estimated bits $\sqcup_j\tilde{\mathbf{s}}_j^{X|Z}$. They are then translated to  
\begin{equation}\label{eq:LUT}
\tilde{E}^{X|Z}=\tilde{\mathcal{L}}_c^{X|Z}\mathscr{T}(\sqcup_j\tilde{\mathbf{s}}_j^{X|Z}) = \tilde{\mathcal{L}}_c^{X|Z}\prod_j\mathscr{T}(\tilde{\mathbf{s}}_j^{X|Z})
\vspace{-3mm}
\end{equation}
through an LUT with $\frac{L^2-1}{2}$ entries recording $\mathcal{L}_c^{X|Z}$ and $\{\mathscr{T}(h_k^{X|Z})\}$, where $h_k$ is an $L^2$-length binary string with all zeros except for the $k$-th bit. Equation~\eqref{eq:LUT} corresponds to a linear combination of these entries, which is a series of pairwise Exclusive-OR (XOR) operations. Afterwards, the error information is transmitted to the control module to generate error correction signals. The total memory consumption for LUTs is $2 \times (\frac{L^2-1}{2}) \times L^2 = L^4 - L^2$ bits. 
Such memory requirements are relatively small and can be easily implemented using LUT for foreseeable code distances (e.g. only 3.5 KB for $L=13$). 
Therefore, the main memory consumption of our NN decoder is determined by the number of network parameters.

\subsection{Network-Specific Architecture} \label{sec:resource_allocation}
Our network-specific architecture is to divide AUs in the NPE into several groups for different network layers. Connections between adjacent network layers are hard-wired, and each network layer will use a separate portion of the computation resources.

\noindent \textbf{Resource Constraints.}
NPE contributes a significant part to the decoding latency. If sufficient AUs exist, the computation of each layer in NPE can be carried out in a single step and is executed fully parallel, resulting in a very low latency. However, this approach comes at a price of considerable computational resource consumption. Although many algorithmic efforts have been made to reduce the arithmetic cost, this level of hardware overhead still makes the overall architecture not practical. The later evaluation shows that even cutting-edge FPGAs are incapable of achieving a fully parallelized $L=5$ NN decoder (see Section~\ref{sec:eval}).

\noindent \textbf{Resource Allocation Model.}
Therefore, the resource allocation of each network layer needs to be carefully customized for optimal performance. To resolve this issue, we use an allocation model to determine resource partitioning.


Suppose there are $C$ AUs, $n_l$ different NN layers, and $M_j$ multiplications operations for the layer $j$. The problem reduces to a constrained optimization to choose a partition $\{C_j\}$
\begin{equation}\label{eq:resource_alloc}
\min_{\{C_j\}}\sum_j^{n_l} \alpha_j \frac{M_j}{C_j}, \quad {\text{subject to}}\quad \sum_j \alpha_j C_j = C
\end{equation}
Here, $\alpha_j$ is the number of independent parts for the layer $j$, which equals to $1$ for the frontend and $>1$ for the backend. This problem can be solved through the Lagrange multiplier, obtaining some real-valued solution $\{C_j\}$, which can be rounded to integers with the equation constraint satisfied. It turns out that this simple heuristics is efficient and exhibits excellent performance in our experiments.

\subsection{Multi-core NPE for Large Distance.}\label{sec:multi-core}

\begin{figure}[h]
\centering
\includegraphics[width=0.47\textwidth]{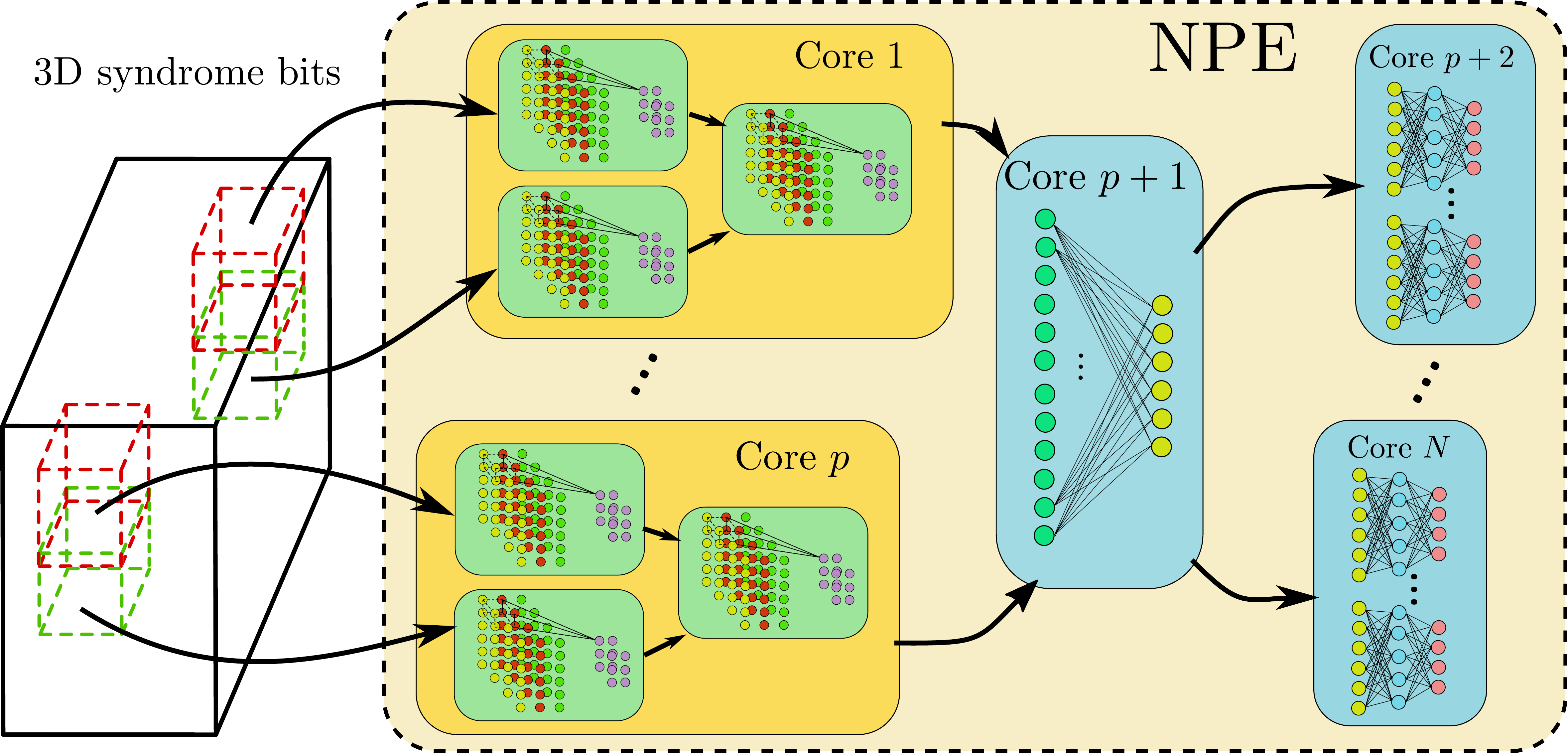}
\setlength{\abovecaptionskip}{3pt}
\captionsetup{labelfont={small,bf}, font={small,bf}}
\caption{A multi-core NPE for large distance $L$.
\label{fig:multi-core}}
\vspace{-3mm}
\end{figure}

Note that the computational complexity grows as $O(L^3)$ (Equation~\eqref{eq:compu.comple}), which puts a hard limit on code distance $L$ with the corresponding decoding algorithm being able to be efficiently executed on a single processing core with constrained computational resource.  The intrinsic parallelism inside MTLND can be exploited to distribute the computation of the NN to a multi-core NPE. A simplified illustration of such approach is shown in Figure ~\ref{fig:multi-core}. The cores form a tree structure, with each core responsible for a part of computation in the 3D CNNs/FCNs. In the context of 3D CNNs with a stepwise structure, the inputs for different cores are approximately independent, necessitating minimal core-to-core communication. It should be noted that this approach is infinitely parallelizable—by fully utilizing each core, the computational scale can be expanded by adding more cores, maintaining a decoding latency of $O(\log L)$. For large-scale FTQEC involving multiple logical qubits decoded using this microarchitecture, syndrome compression as described in~\cite{das2022afs} can also be employed to conserve bandwidth.





\subsection{Exploiting Parallelism in Multi-rounds Measurements}

\begin{figure}[h]
\includegraphics[width=0.48\textwidth]{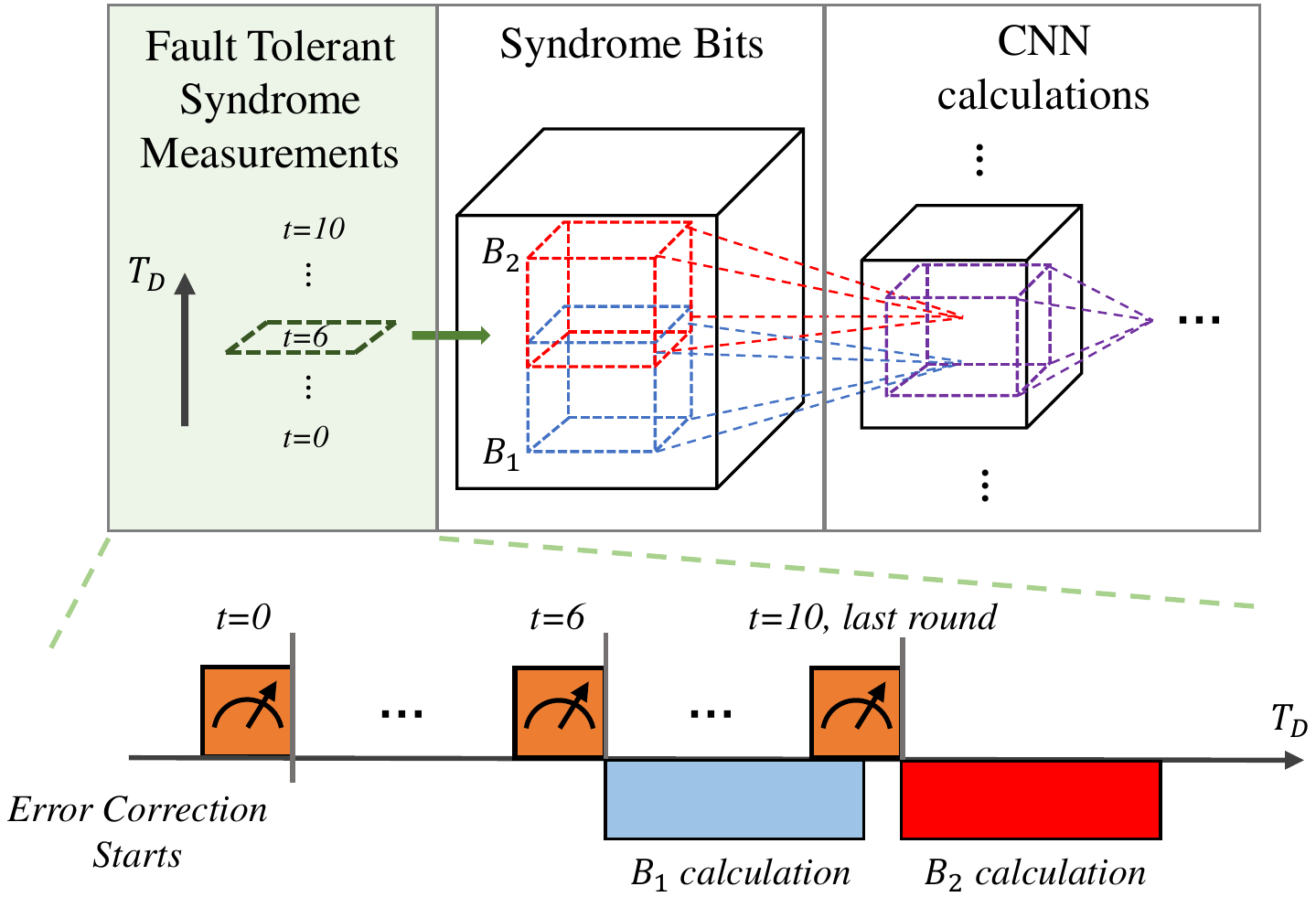}
\centering
\captionsetup{labelfont={small,bf}, font={small,bf}}
\caption{Timeline of sliding window decoding. }
\label{fig:timeline}
\vspace{-3mm}
\end{figure}

The decoupled frontend of MTLND allows independent executions of multiple partitioned input information blocks. The syndrome bits collected from $T$ rounds of measurements form a 3D array input to the NPE, which can be divided into multiple information blocks.
The results of each SM round are independent and arrive at the decoder sequentially in intervals of an SM period. Such features provide certain degree of parallelism that can be exploited---instead of waiting for all syndrome bits to arrive, we prefetch information blocks that are prepared ahead of other blocks, so that different blocks can be processed in a pipeline. An example of such sliding window decoding is shown in Figure \ref{fig:timeline}.

\section{Programmable Decoder}\label{sec:programmable_arch}
In this section, we present an architectural design to support a programmable decoder. This programmable architecture presents better scalability and flexibility compared to the network-specific architecture.


\subsection{Limitations of Network-Specific Architecture}

The network-specific architecture provides good latency performance for small-sized networks due to the customized computational units of each network layer. 
Although many algorithmic efforts are made and comparably low computational complexity is achieved, the resource constraint on this approach is still stringent for large NNs. Therefore, this architecture suffers from limited scalability when scaling to large code distances. 
Furthermore, the implemented decoder is restricted to work for specific NNs, resulting in poor flexibility for different decoder configurations. 
This problem becomes severe when switching to ASICs in the future, which provides better optimized performance but lacks the programmability of FPGA.
Finding a solution providing flexibility while alleviating resource constraints is challenging. Meeting latency requirements further complicates the design, as additional latency overhead is often required to provide flexibility.

\subsection{Insight: Maximizing Resource Utilization within a Given Time Frame}

A single instance of syndrome-array decoding necessitates resource optimization within the decoding duration, which is distinct from the emphasis on high average throughput in conventional NN accelerators. Given the fact that decoders' different layers of networks do not function simultaneously, we employ a \textit{generalized} NPE design adaptable to various NN structures, maximizing resource utilization by allocating all available AUs to each layer, and enhancing scalability for larger code distances with moderate latency impact. Moreover, the generalized NPE enables the development of programmable decoders.

\subsection{Proposal: Programmable Architecture}\label{sec:programmable_micro}

\begin{figure}[t]
\includegraphics[width=0.45\textwidth]{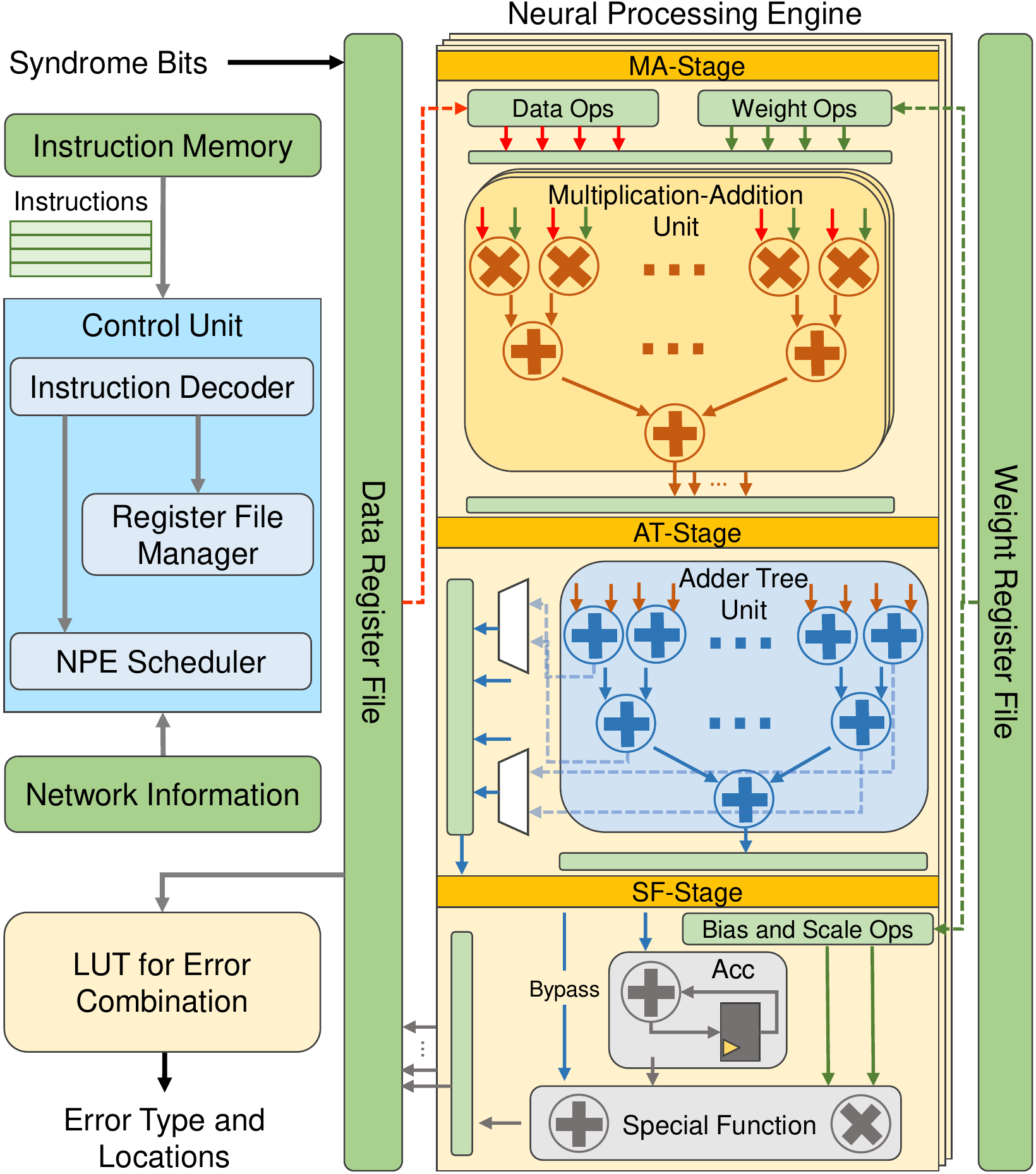}
\centering
\setlength{\abovecaptionskip}{3pt}
\captionsetup{labelfont={small,bf}, font={small,bf}}
\caption{Microarchitecture of the programmable decoder.}
\label{fig:programmable}
\vspace{-5mm}
\end{figure}

We propose a programmable architecture to achieve flexibility and better scalability.
The basic idea is to decompose the execution of each NN layer into a generalized three-stage process and describe it using assembly-level instructions. The decoder microarchitecture is also restructured to accommodate the instruction-based execution.
Designing a dedicated architecture for neural decoders is non-trivial because unlike previously proposed machine learning accelerators \cite{chen2014diannao, ham2021elsa}, the entire framework needs to be tailored to achieve low latency for a single inference task.
In this microarchitecture, we minimize this latency by ruling out unnecessary memory transfers and customizing the control mechanism in the control unit.
It turns out that the gains due to flexibility and resource savings outweigh the latency overhead.
The overview of our proposed microarchitecture of the programmable decoder is shown in Figure \ref{fig:programmable}.

\noindent \textbf{Control Unit.}
Before FTQC begins, a series of assembly codes describing the network structure in the decoding algorithm is generated and loaded into the instruction memory. Instructions fetched from the instruction memory are decoded and then assigned to control the NPE or manage the register file.
Basic network information is also pre-stored in memory and is accessed by the control unit during run-time.
The NPE scheduler receives commands about computations, and determines the specific operations to be performed in the NPE using a finite state machine (FSM). 
The register file manager is responsible for scheduling the communication between registers and the input/output operands collector at each stage of NPE.
The contents of the register files are then used to perform computations at various stages in the NPE.


\noindent \textbf{Three-stage NPE.}
Instead of implementing specific AUs for different layers, we divide the NPE processing into three stages and applied to all AUs. Each stage is customized to the layer types used in our decoding algorithm. This microarchitecture implements multiple processing engines to fit the vector operations, and the following descriptions take one column as an example.
The first stage consists of multiple multiplication-addition units (MAU), which multiply two sets of inputs and add all element-wise products to output the final result.
Multiple parallel MAUs can help us flexibly choose how the mathematical operations of the network layers are constructed. 
This stage completes the primary workload of each layer.

The next stage consists mainly of an adder tree (AT), which has a depth of $\log_2 c$ when there is $c$ MAUs in the MA stage. We can directly connect the output of the MA-stage to the input of the adder tree. A series of multiplexers are used to pre-fetch internal results at different depths within the adder tree, allowing flexible configuration of the MAU operations. Most importantly, this scheme helps reduce decoding latency when only part of the AT is needed for certain layer.
The output of the adder tree is sent to the subsequent special function (SF)-stage, where it is summed with the bias and applied to a scaling factor for activation. The final result is then quantized and written to the data register file, waiting to be fetched as input for the next layer operations.




\noindent \textbf{Single layer divided into multiple chunks.}
A single matrix-vector calculation can be too large to be finished in a single parallel NPE process. Therefore, the input data of this layer is divided into multiple chunks and calculated sequentially based on the scheduling of control instructions. Hence, an accumulator is implemented in the SF-stage to complete the accumulation of the execution results of different chunks. This stage can also be bypassed according to the NPE scheduler. There are also many occasions where multiple layers can be processed in parallel, and prefetching in the AT-stage can help achieve this parallelism.

\noindent \textbf{Control Instructions.}
Compared to classical processors, the error decoding is a static process and the number of NPE execution rounds can be pre-determined based on the network size. Therefore, we can choose the Very Long Instruction Word (VLIW) approach to minimize the instruction execution latency. 
The control instructions for our programmable decoder can be divided into two groups: \textit{computation} and \textit{memory transfer}. These two groups of instructions are used to command the NPE scheduler and register file manager respectively. Hence, the design of control instructions basically represents the method to operate the configurable FSM in the control unit.
The reason for dispatching instructions based on different groups is that we can overlap the latency of reading memory with the time spent on NPE execution, thereby reducing overall latency.
\section{System Implementation}\label{sec:implementation}
In order to give a comprehensive evaluation of our design, we built an FPGA-based system consisting of the decoder itself and control hardware for readout and error correction.

\subsection{Decoder Implementation}
We use Intel Stratix 10 family FPGAs to implement our decoder. We mainly completed two types of implementations:
(1) We first implemented $L=5$ and $L=7$ decoders whose NPE is realized using the \textit{network-specific architecture} as we discussed in \ref{sec:resource_allocation}. 
These implementations are integrated into the evaluation platform to test the performance of near-term error decoding process.
(2) On this basis, we also implemented the microarchitecture of the \textit{programmable decoder} (see \ref{sec:programmable_micro}) to further evaluate the flexibility and scalability of our design.
For all implementations, we focus on implementing NPE with single core.
We use two FPGAs to process the decoding for $X$ and $Z$ errors separately. The subsequent descriptions are given based on one FPGA. 

\noindent \textbf{Network-Specific Implementation (NSI):}
We use $T=10$ syndrome measurement rounds for $L=5$ and $T=14$ rounds for $L=7$. 
These quantities of measurement rounds enable us to assess our architecture's ability to manage large syndrome inputs. Our decoder can readily transition to a smaller number of measurement rounds when practical circumstances permit.
Therefore, The input syndrome results for each error type consist of 120 bits and 336bits, respectively.
We trained different configurations for this design, which determines the memory consumption of the parameter file.
Other minor memory consumption includes registers and flip-flops implemented to store the inputs and outputs during calculations.
These are all implemented using the embedded memory of FPGAs.

The main resource overhead comes from the NPE. 
We prioritize the use of digital signal processing (DSP) units to implement NPE for faster processing. 
All logical Operations are tailored to the constraints of the DSP to fully exploit the limited resource on the FPGAs.
Each round of computation begins by reading new weights into the multiplexer, and the data flow is already hard-wired between different layers.

\noindent \textbf{Programmable Architecture:}
In this implementation, the NPE is structured as a three-stage unit and can be reused by all network layers in the NSI, as well as other different network structures and code distances.
Instead of maximizing the utilization of FPGA resources, we take the the largest layer in the NSI, $\max(C_j)$ in Equation \eqref{eq:resource_alloc}, as the resource constraint for this implementation. This helps us evaluate the effectiveness of our programmable decoder and gain a better understanding of its latency performance.

\subsection{Integrating With Control Hardware}

\begin{figure}[t]
\includegraphics[width=0.45\textwidth]{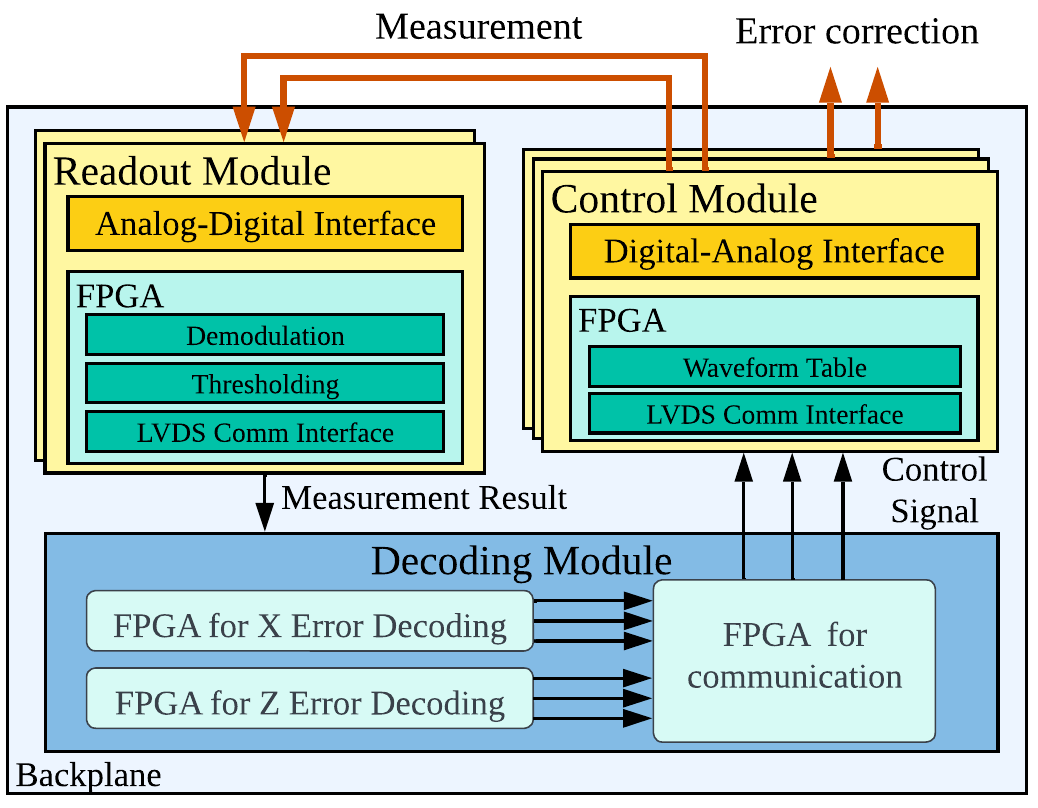}
\centering
\setlength{\abovecaptionskip}{3pt}
\captionsetup{labelfont={small,bf}, font={small,bf}}
\caption{Hardware structure of the implemented decoding system. For evaluation purposes, we connected the measurement signal output of the control module directly to the readout module}
\vspace{-4mm}
\label{fig:system}
\end{figure}

The control hardware of the decoding system is also implemented using custom hardware. The schematic of the entire system is shown in Figure \ref{fig:system}. Each analog-digital interface and its counterparts contain sixteen analog-to-digital converters (ADCs) and digital-to-analog converters (DACs), respectively, for digitizing and generating analog signals.
The decoder takes digitized measurement results as the input syndrome bits, and informs the control module to correct errors. All control and readout modules are connected to the decoding module, and a backplane is implemented to provide wiring of these connections. 

\section{Evaluation Results}\label{sec:eval}

\subsection{Near-term Decoders: $L = 5$ and $L = 7$ With Network-Specific Architecture}
We first use the evaluation platform to test the performance of \textit{Network-Specific decoder}, which implements FTQEC for both $L=5$ and $L=7$.

\noindent \textbf{NN structure:} 
Our $L=5$ decoder has one 3D CNN layer and one FCN layer in the frontend and the backend is composed of 3 two-layer FCNs. The NN structure of $L=7$ decoder is larger: three 3D CNN layers and one FCN layer in the frontend, and 3 two-layer FCNs for the backend. For evaluation, we choose two regimes for the number $N$ of parameters for $L=5$:  $N\approx90$K and $N\approx330$K. $L=7$ decoder has $N\approx960$K parameters.

\noindent \textbf{Hardware complexity:}
The resource utilization of each FPGA in the implemented decoding module is shown in Table \ref{table:complexity}. We used two FPGAs to achieve complete error decoding functionality. Regarding the logic resources, DSP blocks and Adaptive Logic Modules (ALMs) are used for implementing NPE. We utilized these computing resources as much as possible, as discussed in the resource allocation model in Section \ref{sec:resource_allocation}. 
In the implementation of $L=7$, $N \approx 960 $K, the resource utilization of DSP blocks and ALMs is 82\% and 76\%, respectively. A higher level of resource utilization will hamper FPGA routing and can make synthesis fail. The resource consumption of $L=5$, $N \approx 90 $K is much lower and all network layers are maximally parallelized.
The memory consumption of the decoder primarily comes from the parameter file. As shown in Table \ref{table:complexity}, this level of memory consumption is moderate considering modern FPGAs can provide 10-20 MB of embedded memory.

\begin{scriptsize}
\begin{table}[h!]
  \small
  \centering
  \begin{tabular}{|c|c|c|c|}
    \hline
    \multirow{2}{*}{Configuration} & \multirow{2}{*}{Memory Bits} & \multicolumn{2}{c|}{Utilization}\\
    \cline{3-4} & \multirow{2}{*}{} & DSP Block & ALMs \\
    \hline
    \hline
    $L = 5$, $T = 10$, $N \approx 90$K & 114 KB & 21\% & 24\%\\
    \hline
    $L = 5$, $T = 10$, $N \approx 330$K & 532 KB & 81\% & 67\%\\
    \hline
    $L = 7$, $T = 14$, $N \approx 960$K & 1.43 MB & 82\% & 76\%\\
    \hline
  \end{tabular}
  \setlength{\abovecaptionskip}{5pt}
  \captionsetup{labelfont={small,bf}, font={small,bf}}
  \caption{Hardware complexity \vspace{-4mm}}
  \label{table:complexity}
\end{table}
\end{scriptsize}

\begin{figure}[h]
\includegraphics[width=0.45\textwidth]{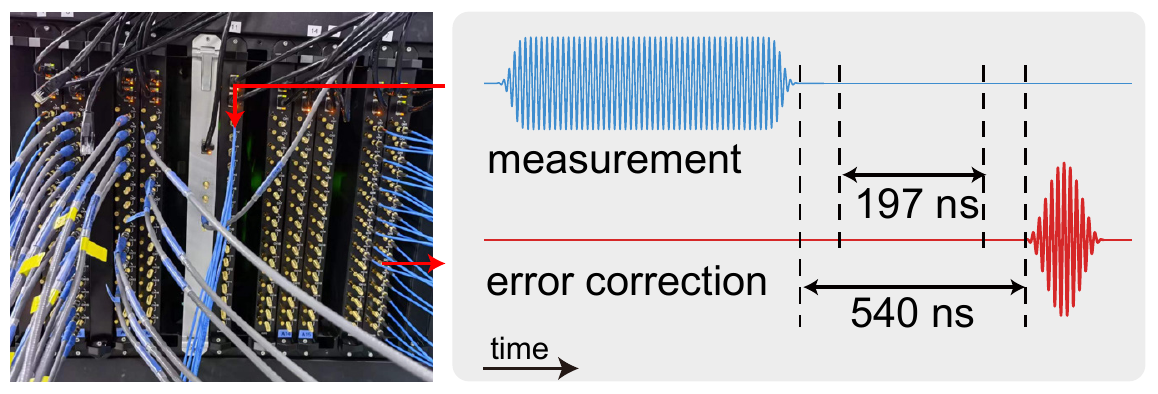}
\centering
\setlength{\abovecaptionskip}{3pt}
\captionsetup{labelfont={small,bf}, font={small,bf}}
\caption{Experimental setup and method for measuring latency.}
\label{fig:latency}
\vspace{-3mm}
\end{figure}

\noindent \textbf{Latency:}
The measured latency results of different configurations are shown in Table \ref{table:latency}. 
The fully-pipelined architecture of the NSI takes 67 cycles to obtain error position for the $L=5$, $N \approx 90 $K configuration, resulting in a decoding latency of \textbf{197 ns}. 
The latency of our $L=7$, $N \approx 960 $K configuration is 1.136 $\mu$s, which is quite good performance considering the resource constraints of current FPGAs. 
Note that this decoding latency is \emph{independent} of the physical error rate $p$. 

\begin{scriptsize}
\begin{table}[h!]
  \small
  \centering
  \begin{tabular}{|c|c|c|c|}
    \hline
    Implementation and & Frequency & Decoding  & Total  \\
    Configuration &  &  Latency &  Latency \\
    \hline
    \hline
    NSI, $L = 5$, $T = 10$, $N \approx 90$K & 330 MHz & 197 ns & 540 ns\\
    \hline
    NSI, $L = 5$, $T = 10$, $N \approx 330$K & 300 MHz & 267 ns & 610 ns\\
    \hline
    NSI, $L = 7$, $T = 14$, $N \approx 960$K & 250 MHz & 1.136 $\mu$s & 1.48 $\mu$s\\
    \hline
  \end{tabular}
  \setlength{\abovecaptionskip}{5pt}
  \captionsetup{labelfont={small,bf}, font={small,bf}}
  \caption{Latency of different configurations}
  \label{table:latency}
  \vspace{-4mm}
\end{table}
\end{scriptsize}

The total latency of our system is obtained by measuring the time interval between receiving measurement signals and issuing correction signals. We connect these two channels to an oscilloscope for testing, as shown in Figure \ref{fig:latency}. 
The total latency is measured to be \textbf{540 ns}, which is fast enough for near-term FTQEC. 
Our solution supports synchronization and data transmission between dozens of modules, and is the fastest real-time FT decoding system ever built for surface code of approximately 100 qubits .

\begin{figure}[ht]
\includegraphics[width=0.45\textwidth]{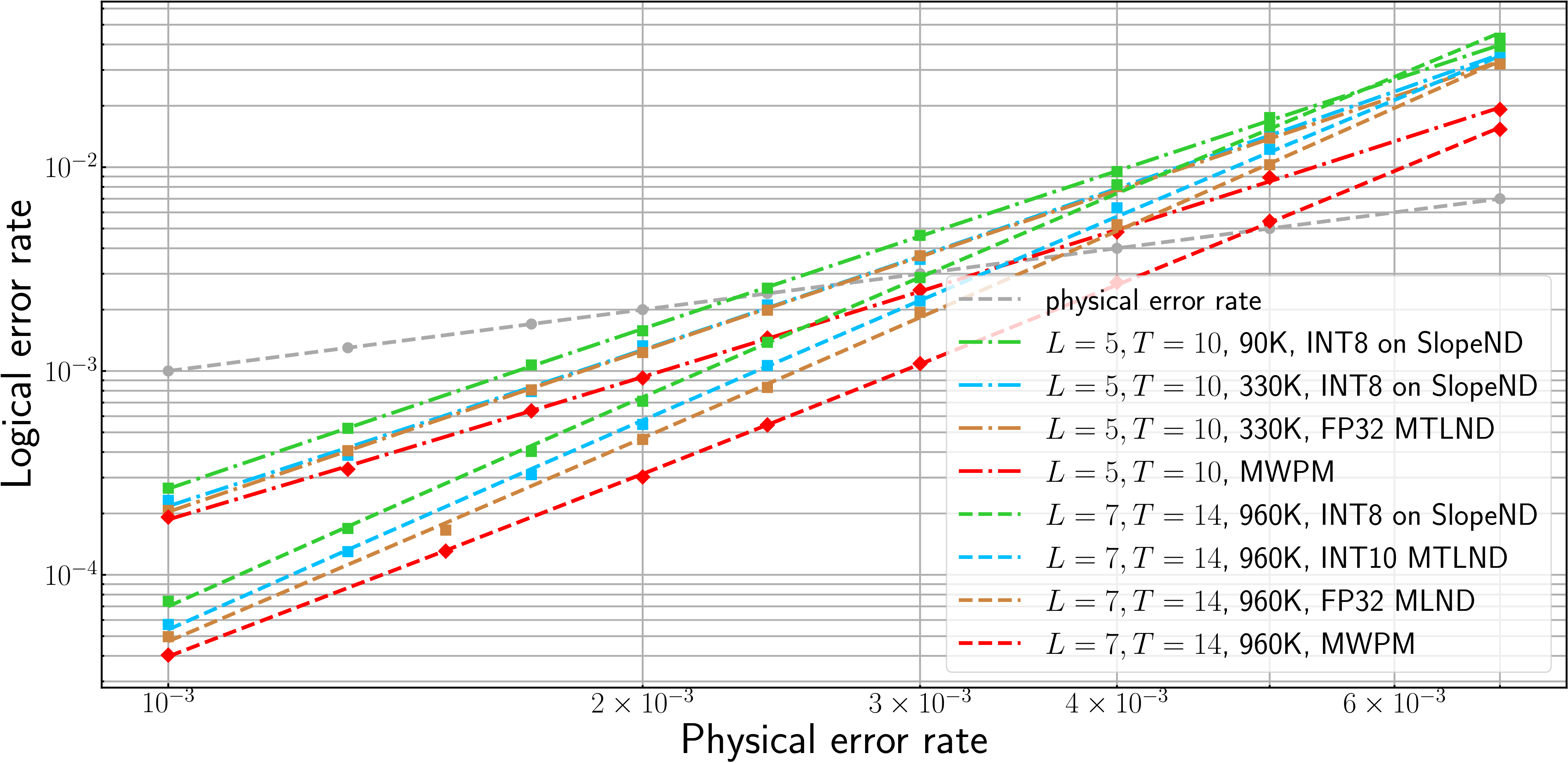}
\centering
\setlength{\abovecaptionskip}{3pt}
\captionsetup{labelfont={small,bf}, font={small,bf}}
\caption{Real-time decoding performance of $L = 5$ and $L = 7$.}
\label{fig:rt_decoding}
\vspace{-5mm}
\end{figure}

\noindent \textbf{Accuracy:}
Figure~\ref{fig:rt_decoding} shows the logical error rate obtained from performing Monte Carlo experiments using our evaluation platform. 
Our system with different parameter numbers and quantization choices all exhibit close accuracy as MWPM, and the quantization of NNs has small effects on the accuracy. 
This shows that our solution, while achieving very low latency, does not sacrifice the accuracy much. We also notice that our system behaves closer to MWPM when the physical error rate gets smaller, which means that our decoder can be more effective as the quantum hardware progresses.

\subsection{Accuracy of Various Code Distances}
Based on our NSI, we further estimated the accuracy of our MTLND for various code distances. The accuracy results of $L=3$, $L=7$, and $L=9$ (with $T=3$, $T=14$, and $T=12$) are also obtained using software simulation. Specifications of these configurations are shown in Table.~\ref{table:Specs}, which shows a moderate scaling, suiting for large-scale FTQEC. It should be noted that for $L = 11$, the MTLND employs both $X$ and $Z$ syndromes with a sufficiently complex NN to showcase its ability to achieve accuracy close to MWPM in larger scale.
\begin{scriptsize}
\begin{table}[h!]
  \small
  \centering
  \begin{tabular}{|c|c|c|c|c|}
    \hline
    Configuration & \#.Layers & \#.Params & \#.Mults & \#. Train. Data  \\
    \hline
    
    \hline
    $L=3$, $T=3$   & 3 & $ \sim$60K & $\sim$2M &  $\sim$2M\\
    \hline
    $L = 5$, $T = 10$ & 4 & $\sim$330K & $\sim$400K & $\sim$10M\\
    \hline
    $L = 7$, $T = 14$ & 6 & $\sim$960K & $\sim$3.17M  & $\sim$100M\\
    \hline
    $L = 9$, $T = 12$ & 8 & $\sim$2.3M & $\sim$10M &  $\sim$240M\\
    \hline
    \hline
    $L = 11$, $T = 11$ & 10 & $\sim 17 $M  & $\sim$ 87M  & $\sim$ 300M\\ 
    \hline
  \end{tabular}
  \setlength{\abovecaptionskip}{5pt}
  \captionsetup{labelfont={small,bf}, font={small,bf}}
  \caption{NNs Specs and resource for MTLND.}
  \label{table:Specs}
  \vspace{-4mm}
\end{table}
\end{scriptsize}

\begin{figure}[h]
\includegraphics[width=0.45\textwidth]{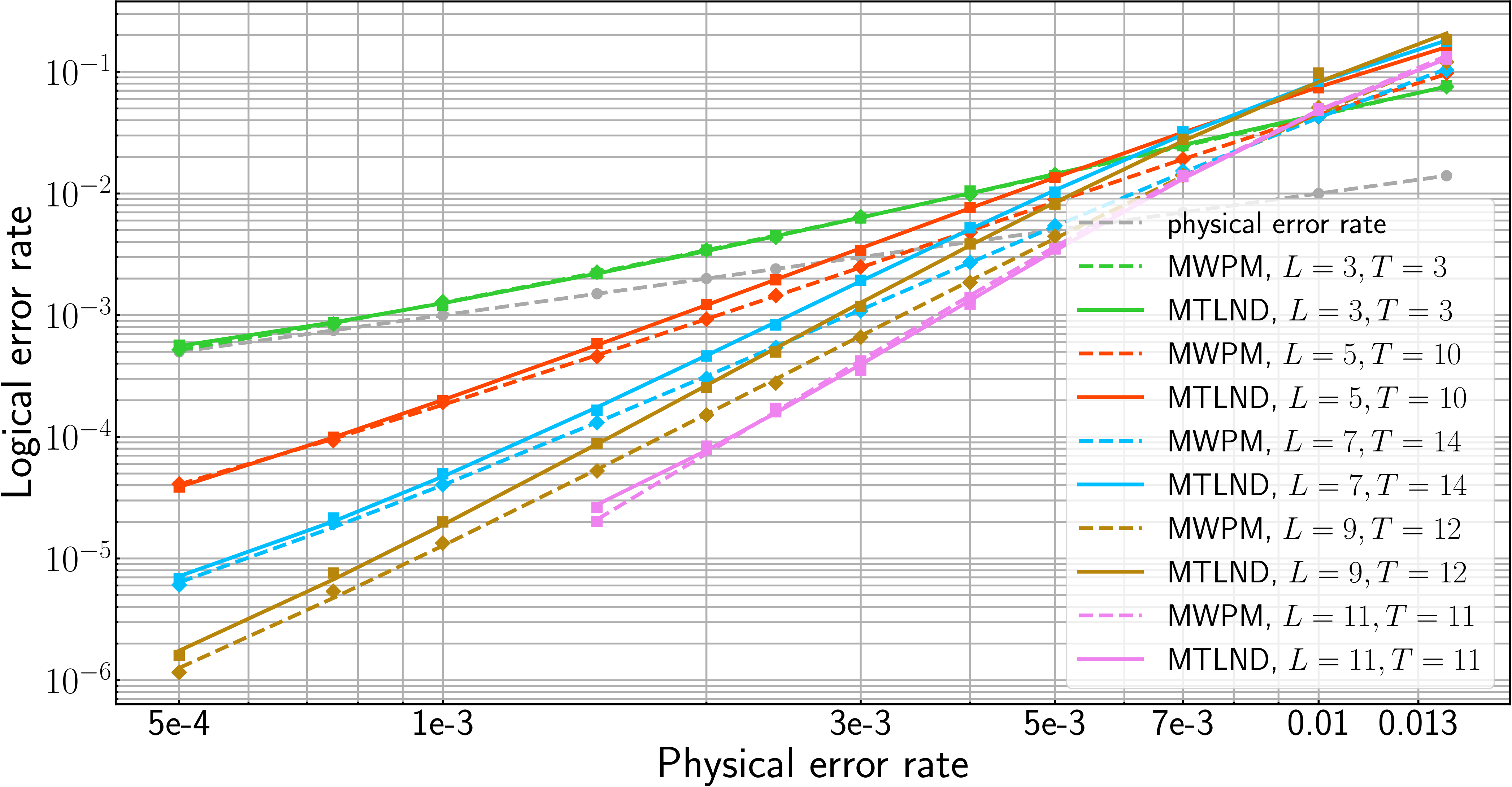}
\centering
\setlength{\abovecaptionskip}{3pt}
\captionsetup{labelfont={small,bf}, font={small,bf}}
\caption{Logical error rate for different code distance.}
\label{fig:logical_error_rate}
\vspace{-4mm}
\end{figure}

Their logical error rates are shown in Figure~\ref{fig:logical_error_rate}, which are all close to their MWPM counterparts while achieving a high accuracy threshold around 0.8\%.

\begin{figure}[h]
\includegraphics[width=0.45\textwidth]{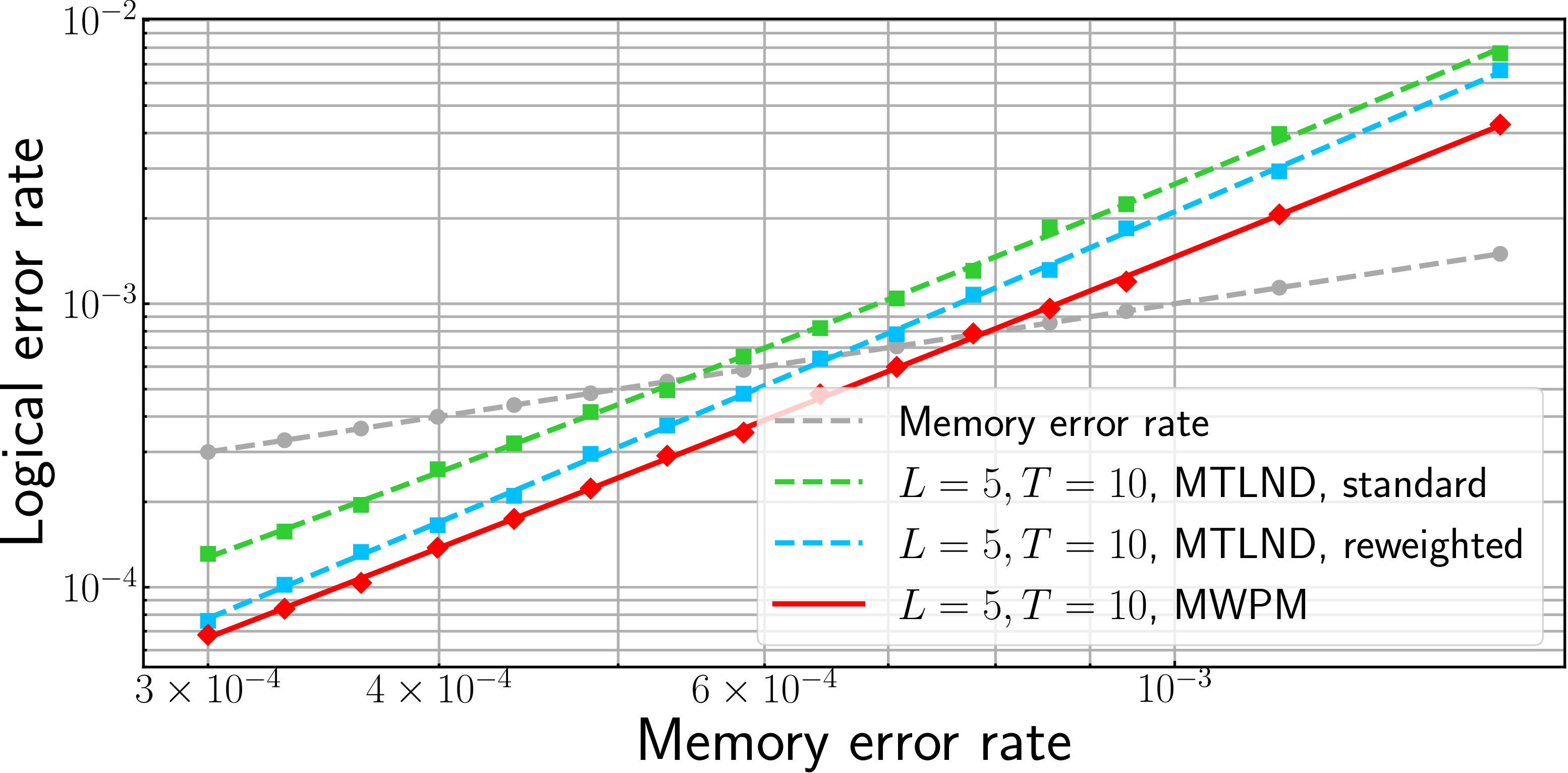}
\centering
\setlength{\abovecaptionskip}{3pt}
\captionsetup{labelfont={small,bf}, font={small,bf}}
\caption{Logical error rate for standard and reweighted MTLND in the case $p_s:p_g:p_m = 1:3:5$}
\label{fig:BM_noise_models}
\vspace{-5mm}
\end{figure}

In actual QEC experiments, one can not access the accurate noise model, which typically differs from the error model used to train the MTLND. Here, we consider the error model when $p_s:p_g:p_m=1:3:5$, which fits the reality that gates and measurement error rates are much larger than the single qubit memory error for superconducting qubits. Figure~\ref{fig:BM_noise_models} shows the logical error rate for the same network trained by the standard training set (standard MTLND) and the one generated by $p_s=0.0024$, $p_g=0.0072$ and $p_m = 0.012$ (reweighted MTLND). This demonstrates that the MTLND can still operate effectively with a slight performance trade-off, while the reweighted version maintains a similar level of accuracy to MWPM.

\subsection{Compared to prior decoders}

\begin{figure}[h]
\includegraphics[width=0.45\textwidth]{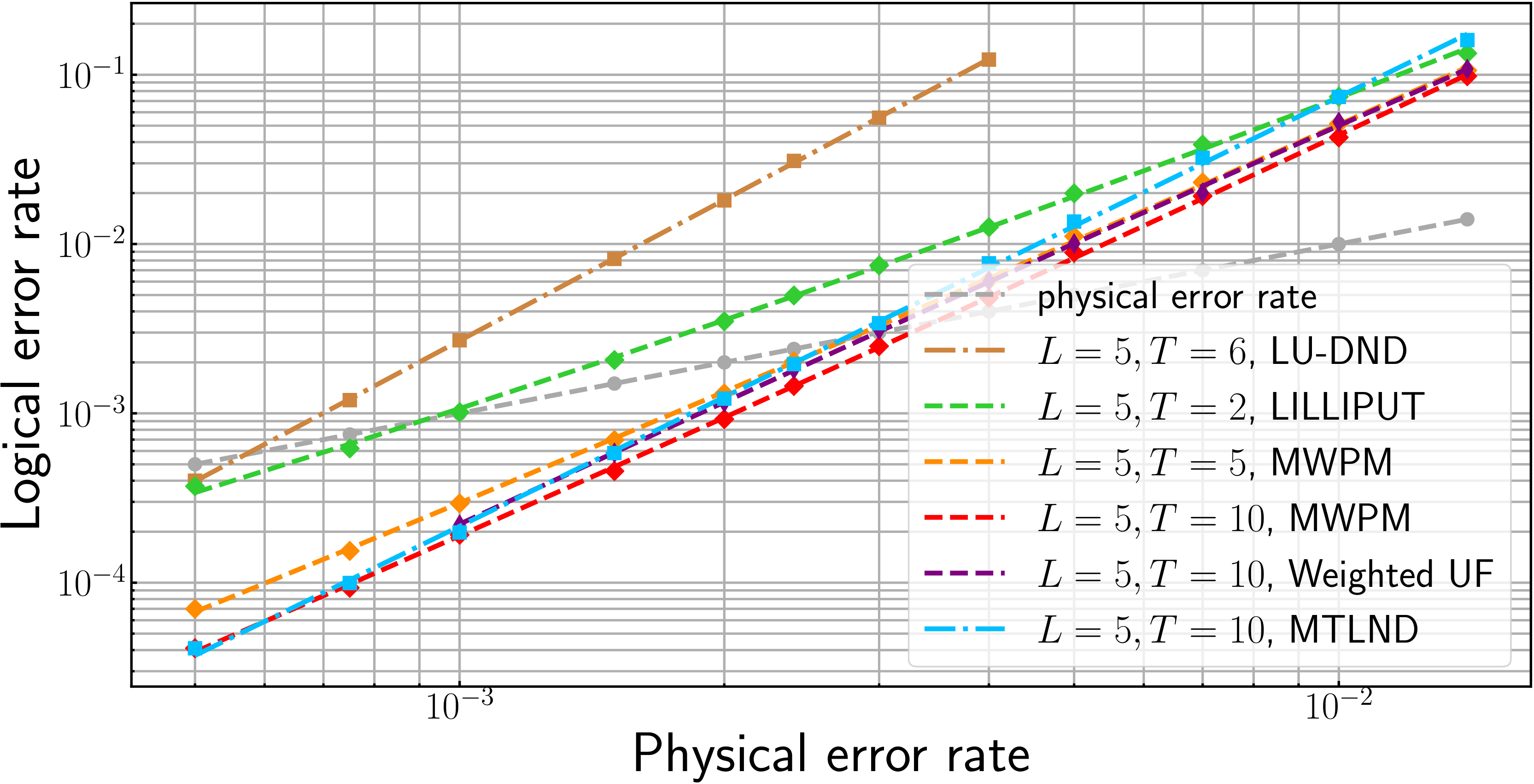}
\centering
\setlength{\abovecaptionskip}{3pt}
\captionsetup{labelfont={small,bf}, font={small,bf}}
\caption{Decoding performance between different decoders for $L=5$}
\label{fig:LE_comparison}
\vspace{-4mm}
\end{figure}

Figure~\ref{fig:LE_comparison} compares MTLND with various decoders proposed. It is clear that the MTLND with $T=10$ outperforms both LU-DND~\cite{chamberland2018deep} and LILLIPUT~\cite{das2021lilliput} and is comparable with weighted UF~\cite{weighted_UF_huang2020}.

\subsection{Programmable Architecture}

\begin{figure}[h]
\includegraphics[width=0.46\textwidth]{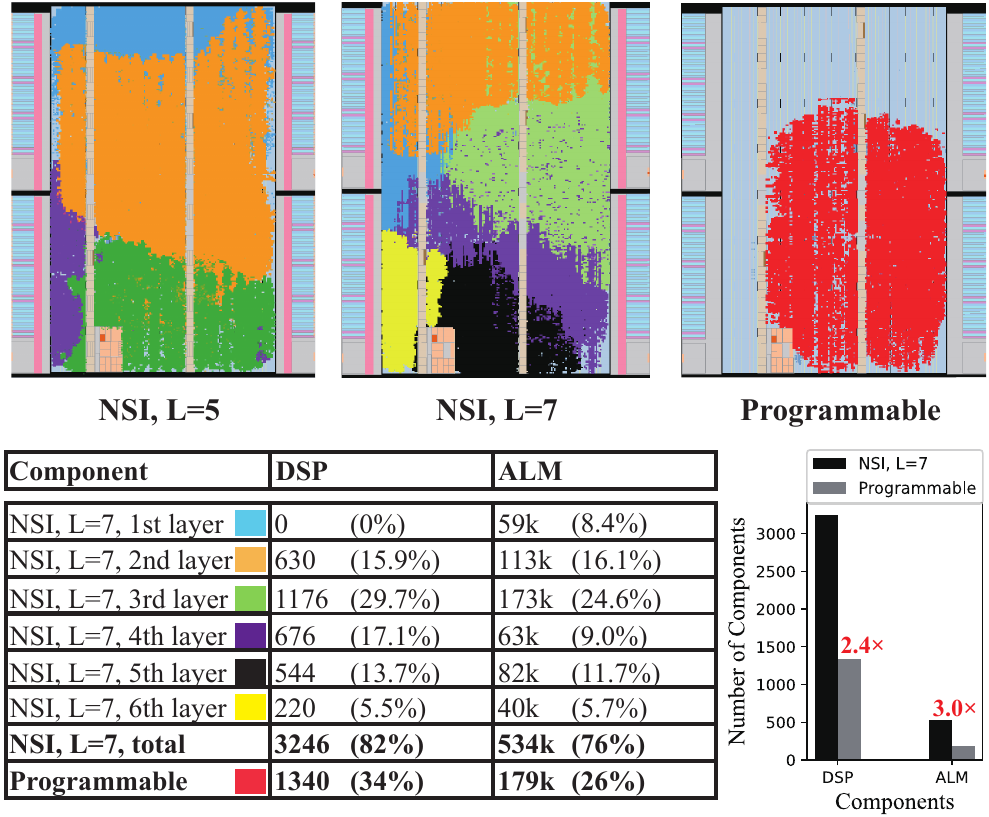}
\centering
\setlength{\abovecaptionskip}{3pt}
\captionsetup{labelfont={small,bf}, font={small,bf}}
\caption{FPGA resource utilization of our NSI ($L=7$, $N \approx 960$K) and programmable decoder.}
\label{fig:prog_result}
\vspace{-5mm}
\end{figure}

\noindent \textbf{Hardware complexity:} The FPGA resource utilization comparison of our programmable decoder and the NSI is shown in Figure \ref{fig:prog_result}. Note that the same set of arithmetic units in the programmable decoder are applied to all network layers. As a result, it achieves a \textbf{2.4$\times$} reduction in DSP blocks and \textbf{3.0$\times$} in ALMs. This result proves that our programmable architecture effectively reduces the resource consumption and presents better scalability.

\noindent \textbf{Reconfigurability and decoding latency:}  
We tested various configurations on the programmable decoder. All these configurations work correctly and have been verified using the evaluation platform. The decoding latency results of these configurations are shown in Table \ref{table:programmable_latency}. Comparing to NSI, our programmable architecture incurs only a small latency loss for substantially reduced resource overhead.  
Note that this programmable decoder is implemented with a small portion of the FPGA computational resource. A fully-utilized programmable decoder can potentially have better latency performance than the corresponding NSI.
Furthermore, we have also tested $L=9$ configuration, proving that our programmable decoder is capable of handling decoders with large code distances.
\begin{scriptsize}
\begin{table}[h!]
  \small
  \centering
  \begin{tabular}{|c|c|c|}
    \hline
    Implementation & Frequency & Decoding  \\
    and Configuration & & Latency \\
    \hline
    \hline
    Programmable, $L = 5$, $T = 10$, $N \approx 90$K & 260 MHz & 373 ns\\
    \hline
    Programmable, $L = 5$, $T = 10$, $N \approx 330$K & 260 MHz & 454 ns\\
    \hline
    Programmable, $L = 7$, $T = 14$, $N \approx 960$K & 260 MHz & 2.13 $\mu$s\\
    \hline
    Programmable, $L = 9$, $T = 12$, $N \approx 2.4$M & 260 MHz & 
    4.827 $\mu$s\\
    \hline
  \end{tabular}
  \setlength{\abovecaptionskip}{5pt}
  \captionsetup{labelfont={small,bf}, font={small,bf}}
  \caption{Latency of processing different configurations on the programmable decoder}
  \label{table:programmable_latency}
  \vspace{-4mm}
\end{table}
\end{scriptsize}

\noindent \textbf{Estimated performance on ASIC:}
By transitioning to an ASIC platform, our system's performance can be enhanced due to increased clock frequency (assuming 2.5 GHz) and elimination of FPGA-induced extra cycles for loading NN parameters. We assess the $L=7$ and $L=9$ configurations on the FPGA implementation, subsequently estimating ASIC latency results, displayed in Table \ref{table:ASIC_latency}.

\begin{scriptsize}
\begin{table}[h!]
  \small
  \centering
  \begin{tabular}{|c|c|c|}
    \hline
    Configuration & Platform and & Estimated \\
      & Assumed Frequency & Decoding Latency \\
    \hline
    \hline
    $L = 7$, $T = 14$, $N \approx 960$K & ASIC, 2.5 GHz & 170 ns \\
    \hline
    $L = 9$, $T = 12$, $N \approx 2.3$M & ASIC, 2.5 GHz & 394 ns \\
    \hline
  \end{tabular}
  \setlength{\abovecaptionskip}{5pt}
  \captionsetup{labelfont={small,bf}, font={small,bf}}
  \caption{Estimated latency of larger code distances on the programmable decoder}
  \label{table:ASIC_latency}
  \vspace{-6mm}
\end{table}
\end{scriptsize}

\subsection{Test on Google's Experiment Setting}
\begin{figure}[h]
\includegraphics[width=0.4\textwidth]{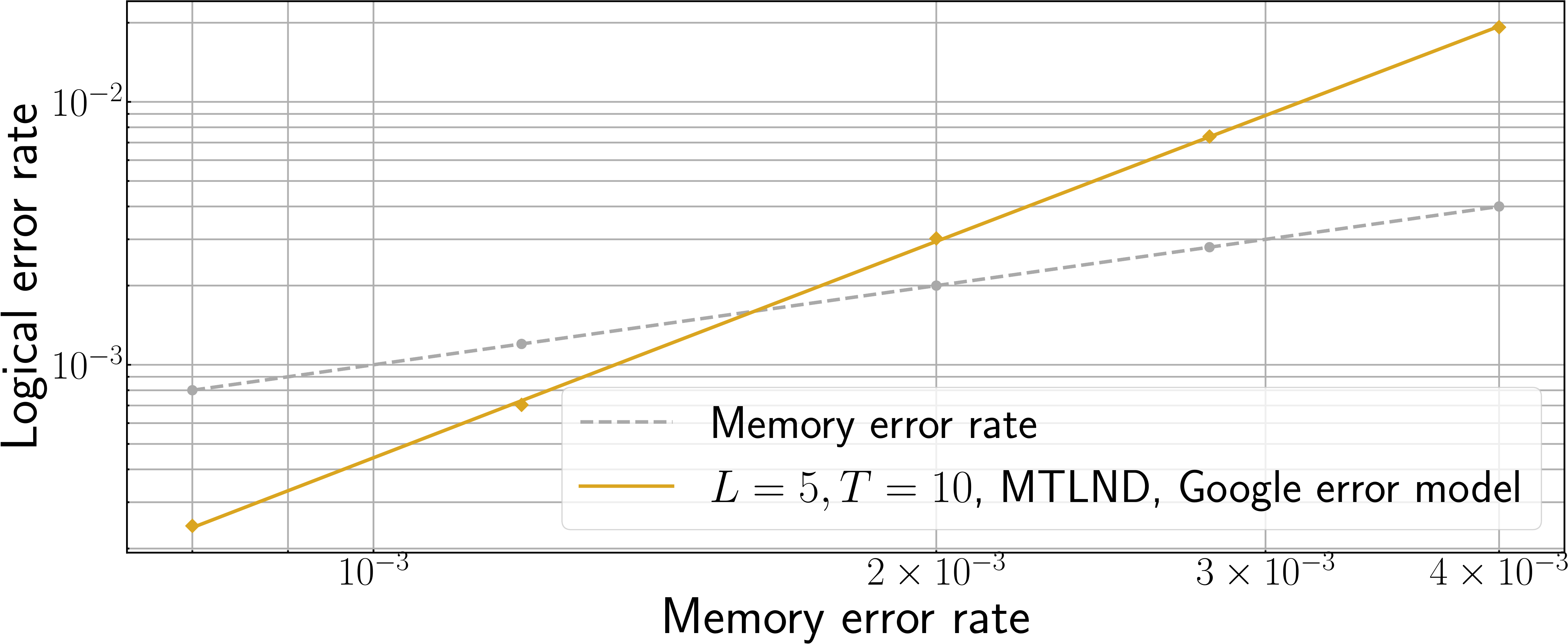}
\centering
\setlength{\abovecaptionskip}{3pt}
\captionsetup{labelfont={small,bf}, font={small,bf}}
\caption{Evaluation of accuracy for the MTLND approach utilizing an error model extracted from experiments conducted by Google.}
\label{fig:LE_google}
\vspace{-7mm}
\end{figure}

We additionally refined our noise model to integrate an effective circuit-level noise representation, informed by Google's experimental data on surface code~\cite{google2023suppressing, Google_data}, with $p_g\sim0.005$, $p_s\sim0.004$, and $p_m\sim 0.018$. The MTLND was trained and assessed under these conditions. Figure.\ref{fig:LE_google} illustrates the accuracy results upon extrapolation to lower noise rates.

\section{Related Work}


The challenges and prospects of real-time decoder research were recently reviewed \cite{battistel2023real}. The review highlights the goal of recent search is to provide concrete evidence that real-time decoding is achievable \emph{in practice}. Our work aims to accomplish this by employing realistic noise models and implementing a comprehensive system.

\noindent \textbf{LUT Decoders.}
The decoder in ~\cite{das2021lilliput} employs an LUT indexed by syndrome bits for error correction search, providing inherent programmability and low latency due to only requiring memory access time. However, this LUT method is not scalable as the number of entries grows exponentially.

\noindent \textbf{Union-Find Decoder \cite{das2022afs, liyanage2023scalable}.}
The UF algorithm potentially offers hardware implementation simplicity, yet parallelizing this graph-based approach for low latency remains challenging. Moreover, in \cite{das2022afs, liyanage2023scalable}, only the phenomenological noise model is considered, while incorporating circuit-level noise would considerably impede the decoder's speed.

\noindent \textbf{Other Neural Decoders.}
In \cite{overwater2022neural}, the networks are restricted to FCNs, limiting their ability to manage large code distances and realistic error models. Chamberland \textit{et al.}~\cite{chamberland2018deep,chamberland2022techniques} investigated CNNs and estimated hardware performance; however, they either exhibited high latency (over 2000 $\mu$s) or unsatisfactory accuracy. To the best of our knowledge, reconfigurable neural decoders have not been previously explored. Furthermore, our programmable solution's architectural benefits enable improved scalability compared to prior work.

\noindent \textbf{SFQ-based Decoders.}
Superconducting Single Flux Quantum (SFQ) technology offers high clock speeds and qubit integration capabilities. However, current SFQ-based decoders~\cite{holmes2020nisq+, ueno2021qecool,ueno2022qulatis,ueno2022neo,ravi2022have} are hindered by limited computational power, resulting in poor accuracy. Scaling up this approach presents a considerable challenge, barring near-term advancements in superconducting logic device densities. 

\noindent \textbf{Real-time QEC Experiments.} Experiments on real-time QECs emerge in past years, including those using the  repetition code~\cite{BBNriste2020real}, Gottesman-Kitaev-Preskill (GKP) code~\cite{rt_GKP} and the distance-3 color code~\cite{ryan2021realization}. Such simple codes are inadequate for handling general or complex noises. Consequently, they are restricted to small-sized QECCs.


\section{Conclusions}

Developing scalable and accurate real-time decoders for FTQEC has been an active area of research. In this work, we propose a neural decoding system, which suits both near-term and large-scale FTQCs. We carry out both algorithmic and architectural optimizations for accuracy, scalability, and low latency. Furthermore, our programmable architecture provides flexibility to explore different decoding configurations to adapt to a variety of FTQEC scenarios. Finally, we build a comprehensive decoding system using off-the-shelf FPGAs to evaluate our design. A demonstration of $L=5$, $T=10$ decoder costs 197 ns on the real device while approaching the comparable accuracy with MWMP under circuit-level noises. The evaluation shows the capability of our system for near-term and large-scale real-time FTQEC.

\section*{ACKNOWLEDGMENTS}
We thank all members of Tencent Quantum Labrotory who contributed to the experimental set-up. This work is funded in part by Key-Area Research and Development Program of Guangdong Province, under grant 2020B0303030002.

\bibliographystyle{IEEEtranS}
\bibliography{refs}

\end{document}